\documentclass[aps,pre,onecolumn,longbibliography,nofootinbib,showkeys,amssymb,floatfix]{revtex4-1}

\usepackage{latexsym}
\usepackage{amsfonts}
\usepackage{color}
\usepackage{graphicx}
\usepackage{mathptmx}      
\usepackage{bm}
\usepackage{enumerate}
\usepackage{subfigure}
\usepackage{listings}
\usepackage{grffile}
\usepackage{multirow}

\newcommand\T{t}
\newcommand\R{r}

\begin{document}
\title{Long-time correlations in single-neutron interferometry data}

\author{M. Willsch}
\affiliation{Institute for Advanced Simulation, J\"ulich Supercomputing Centre,\\
Forschungszentrum J\"ulich, D-52425 J\"ulich, Germany}
\affiliation{RWTH Aachen University, D-52056 Aachen, Germany}
\author{D. Willsch}
\affiliation{Institute for Advanced Simulation, J\"ulich Supercomputing Centre,\\
Forschungszentrum J\"ulich, D-52425 J\"ulich, Germany}
\affiliation{RWTH Aachen University, D-52056 Aachen, Germany}
\author{F. Jin}
\affiliation{Institute for Advanced Simulation, J\"ulich Supercomputing Centre,\\
Forschungszentrum J\"ulich, D-52425 J\"ulich, Germany}
\author{K. Michielsen}
\affiliation{Institute for Advanced Simulation, J\"ulich Supercomputing Centre,\\
Forschungszentrum J\"ulich, D-52425 J\"ulich, Germany}
\affiliation{RWTH Aachen University, D-52056 Aachen, Germany}
\author{T. Denkmayr, S. Sponar}
\affiliation{{\color{black}Atominstitut, Vienna University of Technology\\
A-1020 Vienna, Austria}}
\author{{\color{black}Y. Hasegawa}}
\affiliation{{\color{black}Atominstitut, Vienna University of Technology\\
A-1020 Vienna, Austria}}
\affiliation{{\color{black}Division of Applied Physics, Hokkaido University\\
 Kita-ku, Sapporo 060-8628, Japan}}
\author{H. De Raedt}
\email{deraedthans@gmail.com}
\thanks{Corresponding author}
\affiliation{Institute for Advanced Simulation, J\"ulich Supercomputing Centre,\\
Forschungszentrum J\"ulich, D-52425 J\"ulich, Germany}
\affiliation{Zernike Institute for Advanced Materials,\\
University of Groningen, Nijenborgh 4, NL-9747AG Groningen, The Netherlands}

\date{\today}

\begin{abstract}
{\color{black}
We present a detailed analysis of the time series
of time-stamped neutron counts obtained by single-neutron interferometry.
The neutron counting statistics display the usual Poissonian behavior,
but the variance of the neutron counts does not.
Instead, the variance is found to exhibit a dependence on the phase-shifter setting
which can be explained by a probabilistic model that accounts for fluctuations of the phase shift.
The time series of the detection events exhibit long-time correlations with
amplitudes that also depend on the phase-shifter setting.
These correlations appear as damped oscillations with a period of about $2.8\;\mathrm{s}$.
By simulation, we show that the correlations of the time differences observed in the experiment
can be reproduced by assuming that, for a fixed setting of the phase shifter,
the phase shift experienced by the neutrons varies periodically in time with a period of
$2.8\,\mathrm{s}$.
The same simulations also reproduce the behavior of the variance.
Our analysis of the experimental data suggests that time-stamped data of
single-particle interference experiments may exhibit transient features
that require a description in terms of non-stationary processes,
going beyond the standard quantum model of independent random events.
}

\end{abstract}


\maketitle

\section{Introduction}\label{section1}

In the description of single-neutron interference experiments it is taken for granted
that the detection events can be modeled as independent random variables, meaning that
the number of counts registered within a sufficiently large time interval obeys a Poisson distribution~\cite{RAUC15}.
Accordingly, the conditional probability of detecting a neutron within
a time interval $[t,t+\tau]$ when a previous one was detected at time $t$
is given by the exponential distribution $P(\tau)=\exp(-\tau/\Delta t)/\Delta t$
where $\Delta t$ is the average time between two successive detection events at the same detector~\cite{RAUC15,GRIM01}.
The analysis of neutron-counting statistics of interferometry experiments strongly supports
the hypothesis that the distribution is Poissonian~\cite{RAUC15}.
However, counting statistics only probe the first few  moments of the probability distribution and are
therefore not well-suited to uncover possible correlations in the time series.

In this paper, we analyze the time series of time-stamped neutron counts obtained by single-neutron interferometry
experiments.
The main result of this analysis is that the time series of the counts exhibit unexpected long-time correlations with
amplitudes that change with the phase shifter setting.

The paper is organized as follows.
Section~\ref{section2} describes the experiment and the raw experimental data.
Results of the statistical analysis of the observed neutron counts and their variances are presented in Sec.~\ref{section3}.
We show that the neutron counts display the expected sinusoidal behavior but that
a more complicated statistical model is required to account for the
changes of their variances as a function of the phase shifter setting.
In Sec.~\ref{section4}, we analyze the time series of the time-stamped detection events
in terms of the correlations of time differences and the type of detection events.
We demonstrate that the time series exhibits correlations in the form of damped oscillations
with a period of about {\color{black}2.8} seconds, which is a fairly long time relative to the average
time of 1.3 milliseconds between the detection of two successive neutrons.
Section~\ref{section5} scrutinizes the hypothesis that these correlations are caused
by a time-dependent phase shift by performing simulations using two different models.
Both models are shown to reproduce the long-time correlations observed experimentally.
The paper concludes with the discussion given in Sec.~\ref{section6}.

\section{Neutron interference experiment}\label{section2}

\begin{figure}[t]
\begin{center}
\includegraphics[width=0.95\hsize]{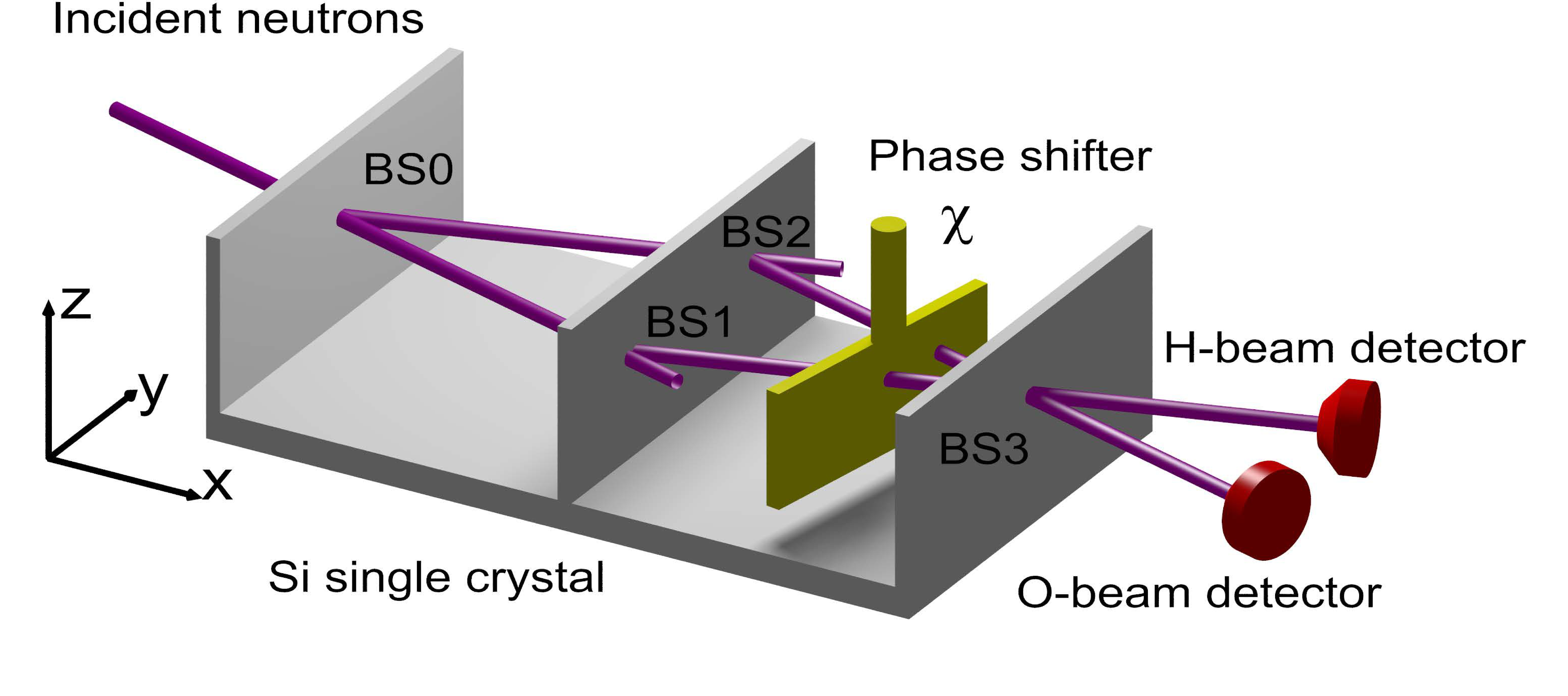} 
\caption{(color online) %
Layout of the perfect Si-crystal neutron interferometer~\cite{RAUC74a}.
BS0, ..., BS3: beam splitters;
phase shifter: aluminum plate;
neutrons that are transmitted by BS1 or BS2 leave the interferometer
and do not contribute to the interference signal.
Detectors count the number of neutrons in the O- and H-beam.
}
\label{fig1}
\end{center}
\end{figure}

\subsection{Description of the experiment}\label{section2a}

Figure~\ref{fig1} shows the setup of the neutron interference experiment.
The three crystal plates, named splitter, mirror and analyzer plate,
are assumed to be identical, meaning that they have the same transmissivity and
reflectivity~\cite{RAUC15}. The three crystal plates have to be parallel to high
accuracy~\cite{RAUC74a} and the whole device needs to be protected from vibrations in order to observe
interference~\cite{KROU00}.
A monoenergetic neutron beam is split by the splitter plate (BS0).
Neutrons refracted by beam splitters BS1 and BS2 (mirror plate) are directed to the analyzer plate (BS3),
also acting as a BS, thereby first passing through a rotatable-plate
phase shifter (e.~g. aluminum plate~\cite{RAUC15}).
Absorption of neutrons by the aluminum plate is assumed to be negligible~\cite{RAUC15}.
Minute rotations of the plate about an axis perpendicular to the base plane of the interferometer
induce large variations in the phase difference~\cite{RAUC15,LEMM10}.
Finally, the neutrons are detected by one of the two detectors placed in the
so-called O-beam or H-beam.
Neutrons which are not refracted by BS1 and BS2 leave the interferometer and
do not contribute to the interference signal.
Neutron detectors can have a very high detection efficiency of almost 100\%~\cite{RAUC15}.
A special feature of the experiments
{\color{black}about which we report in this paper}
is that each detected neutron receives a time stamp.
The records of the time stamps of the neutrons detected in the O- or H-beam
constitute the raw data produced by the experiment.
{\color{black} It has already been demonstrated that the time-stamp data can be used to enhance
the contrast of the interference fringes and phase resolution~\cite{HASE94}.
At that time, the experiments were performed with the low-flux reactor
in Vienna, yielding neutron-counting rates of the order of 1 count per second~\cite{HASE94}.
In contrast, the data analyzed in this paper have been obtained
with the reactor in Grenoble, yielding neutron-counting rates of the order of 1000 counts per second (see below).}

\subsection{Description of the data}\label{section2b}

The experimental data from 37 repetitions of the experiment were collected on 30 April 2015 at the
{\color{black}ILL High-Flux Reactor.}
The procedure is as follows
\begin{enumerate}
\item
With $n=1,\ldots,37$, do steps 2 -- 5.
\item
Reset the phase shifter to its initial position $X=1$.
This position will be called ``setting'' of the phase shifter.
This setting takes values $X=1,\ldots,33$.
\item
While the phase shifter moves from one setting to the next one,
whenever the detector O(H) clicks, write `-' followed by the clock time (= time stamp)
to the O(H) stamp file of run number $n$.
\item
When the phase shifter is in position $X$, count neutrons for 10 seconds and
whenever the detector O(H) clicks, write `+' followed by the clock time (= time stamp)
to the O(H) stamp file of run number $n$.
\item
Increment $X$ and change the setting of the phase shifter accordingly.
Repeat steps 3 to 5 until $X$ exceeds 33.
\end{enumerate}

This procedure, which took about 6 hours and 15 minutes,
yielded 37 pairs of data files containing time-stamps of 9355720
neutrons detected during the time intervals in which the phase shifter stood still.
A very short excerpt from one of these pairs of data files is shown in Table~\ref{tab1}.

\begin{table}[ht]
\caption{
Excerpt of typical time-stamp data, taken from the files ifg\_30Apr1806H.stamp and ifg\_30Apr1806O.stamp
provided to us {\color{black} by the Vienna neutron interferometer group}.
Column one (two) contains the times at which the detector in the O (H) beam registered the arrival of a neutron.
Negative time stamps indicate that the phase shifter is moving from one setting to another one.
The fourth to sixth column show the merged time-stamp data $t_i$,
the time differences $\Delta t_i=t_i-t_{i-1}$  when $t_{i}$ and $t_{i-1}$ are both positive,
and the encoding of the type of event ($x_i=-1(+1),\ldots$ indicating the detection of a neutron in the O (H)-beam), respectively.
Times are in units of $25\;\mu\mathrm{s}$,
and the time interval that the phase shifter is not moving from one setting to another is 10 seconds.
}
\begin{ruledtabular}
\begin{tabular*}{\textwidth}{@{\extracolsep{\fill}} rrc|rrc}
O-beam detector\hfil  & \hfil H-beam detector \hfil & Comment \hfil & \hfil Merged times series ($t_i$)\hfil&\hfil time differences ($\Delta t_i$)\hfil& \hfil Type of event($x_i$)\\
\hline
-1022655    &             &     phase shifter is moving      &  -1022655  &  & $-1$ \\
            &   -1115236  &                                  &  -1115236  &  & $+1$ \\
-1124677    &             &                                  &  -1124677  &  & $-1$ \\
-1139548    &             &                                  &  -1139548  &  & $-1$ \\
+1166969    &             &   phase shifter stopped moving   &  +1166969  &  & $-1$ \\
            &   +1219762  &                                  &  +1219762  & 52793   & $+1$ \\
            &   +1225380  &                                  &  +1225380  & 5618    & $+1$ \\
            &   +1262643  &                                  &  +1262643  & 37263   & $+1$ \\
            &   +1292191  &                                  &  +1292191  & 29548   & $+1$ \\
            &   +1408944  &                                  &  +1408944  & 116753  & $+1$ \\
            &   +1422582  &                                  &  +1422582  & 13638   & $+1$ \\
            &   +1456330  &                                  &  +1456330  & 33748   & $+1$ \\
            &   +1457528  &                                  &  +1457528  & 1198    & $+1$ \\
+1468626    &             &                                  &  +1468626  & 11098   & $-1$ \\
+1559779    &             &                                  &  +1559779  & 91153   & $-1$ \\
            &   +1615907  &                                  &  +1615907  & 56128   & $+1$ \\
            &   +1641410  &                                  &  +1641410  & 25503   & $+1$ \\
            &   +1649128  &                                  &  +1649128  & 7718    & $+1$ \\
+1668761    &             &                                  &  +1668761  & 19633   & $-1$ \\
+1720042    &             &                                  &  +1720042  & 51281   & $-1$ \\
+1779040    &             &                                  &  +1779040  & 58998   & $-1$ \\
  ...       &     ...     &                                  &    ...     &  & ...  \\
            &  +400624979 &                                  & +400624979 &  & $+1$ \\
+400640127  &             &                                  & +400640127 & 15148 & $-1$ \\
            &  +400727895 &                                  & +400727895 & 87768 & $+1$ \\
            &  +400744948 &                                  & +400744948 & 17053 & $+1$ \\
+400815186  &             &                                  & +400815186 & 70238 & $-1$ \\
            &  +400864999 &                                  & +400864999 & 49813 & $+1$ \\
            &  +400947602 &                                  & +400947602 & 82603 & $+1$ \\
            &  -401110120 &   phase shifter is moving again  & -401110120 &  & $+1$ \\
  ...       &     ...     &                                  &    ...     &  & ...  \\
\end{tabular*}
\end{ruledtabular}
\label{tab1}
\end{table}

The first step in processing the time-stamp data is to merge the time stamp data of the O- and H-beam
detectors in one array and attach a label to each element of this array
to keep track of the type of the detection event that the time stamp represents, see Table~\ref{tab1}.

On very rare occasions (642 out of the 9355720 neutrons detected during the time intervals
in which the phase shifter stands still)
we find that there is both an O- and H-event with the same time stamp.
We discard such events in the analysis that follows.

\section{Data analysis}\label{section3}

\subsection{Fitting the neutron counts to a simple quantum model}\label{section3a}

\begin{figure}[t]
\begin{center}
\includegraphics[width=0.80\hsize]{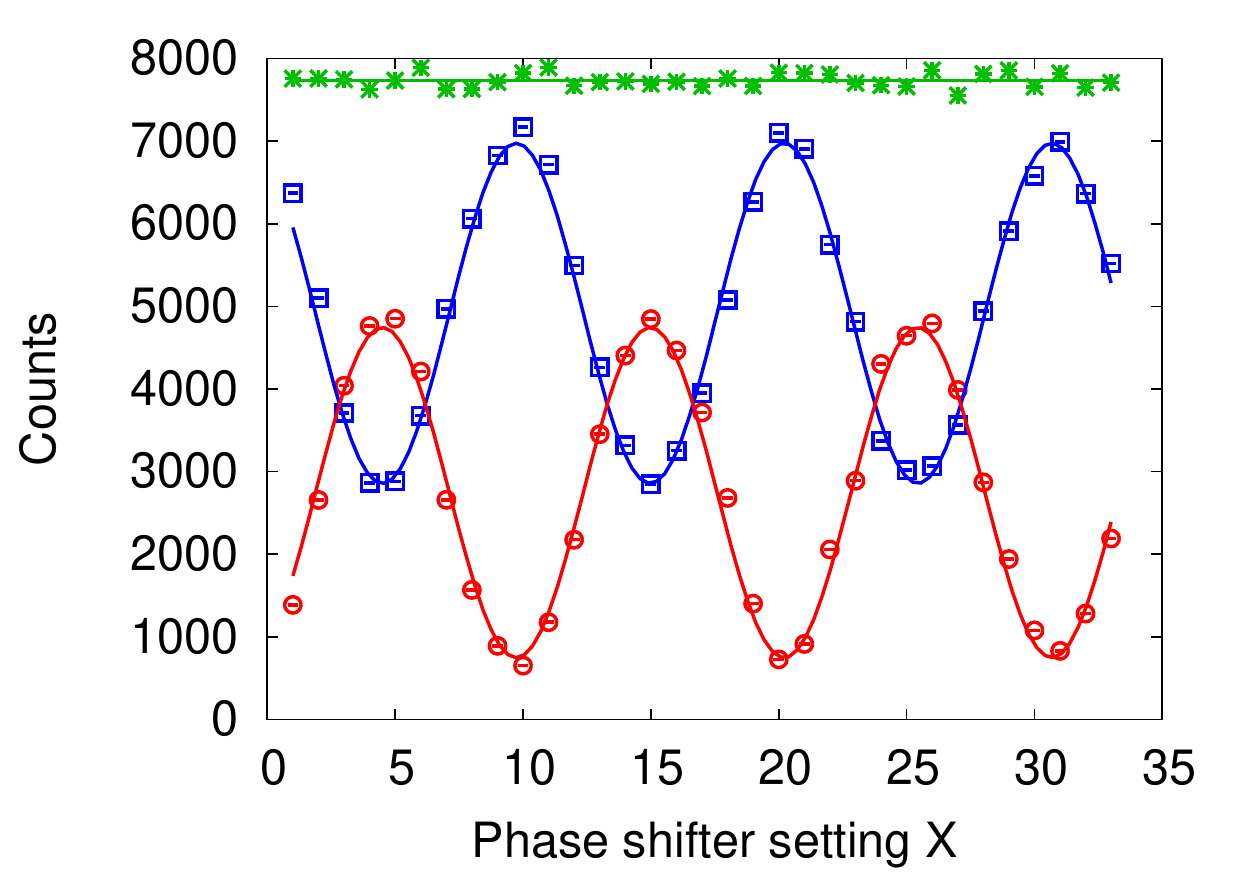} 
\caption{(color online) %
Neutron counts as a function of the setting $X$, as obtained from the first pair of data files
of the experiment.
Open circles ($\bigcirc$): O-beam;
open squares ($\Box$): H-beam;
asterisks ($\ast$): sum of O- and H-beam counts.
Lines through the data points are obtained by fitting to the data as described in the text.
}
\label{fig2a}
\end{center}
\end{figure}

In Fig.~\ref{fig2a}, we present a representative example
of the neutron counts in the O- and H-beam as a function of the setting $X$ of the phase shifter.
Our results for the neutron counts match those found in the data files provided to us
{\color{black}by the Vienna neutron interferometer group.}
As the sum of neutrons detected in the O- and H-beam is nearly constant,
the sinusoidal dependence of the O- and H-beam counts on $X$
provides direct evidence for single-neutron interference~\cite{RAUC74a,RAUC15}.

\begin{figure}[t]
\begin{center}
\includegraphics[width=0.80\hsize]{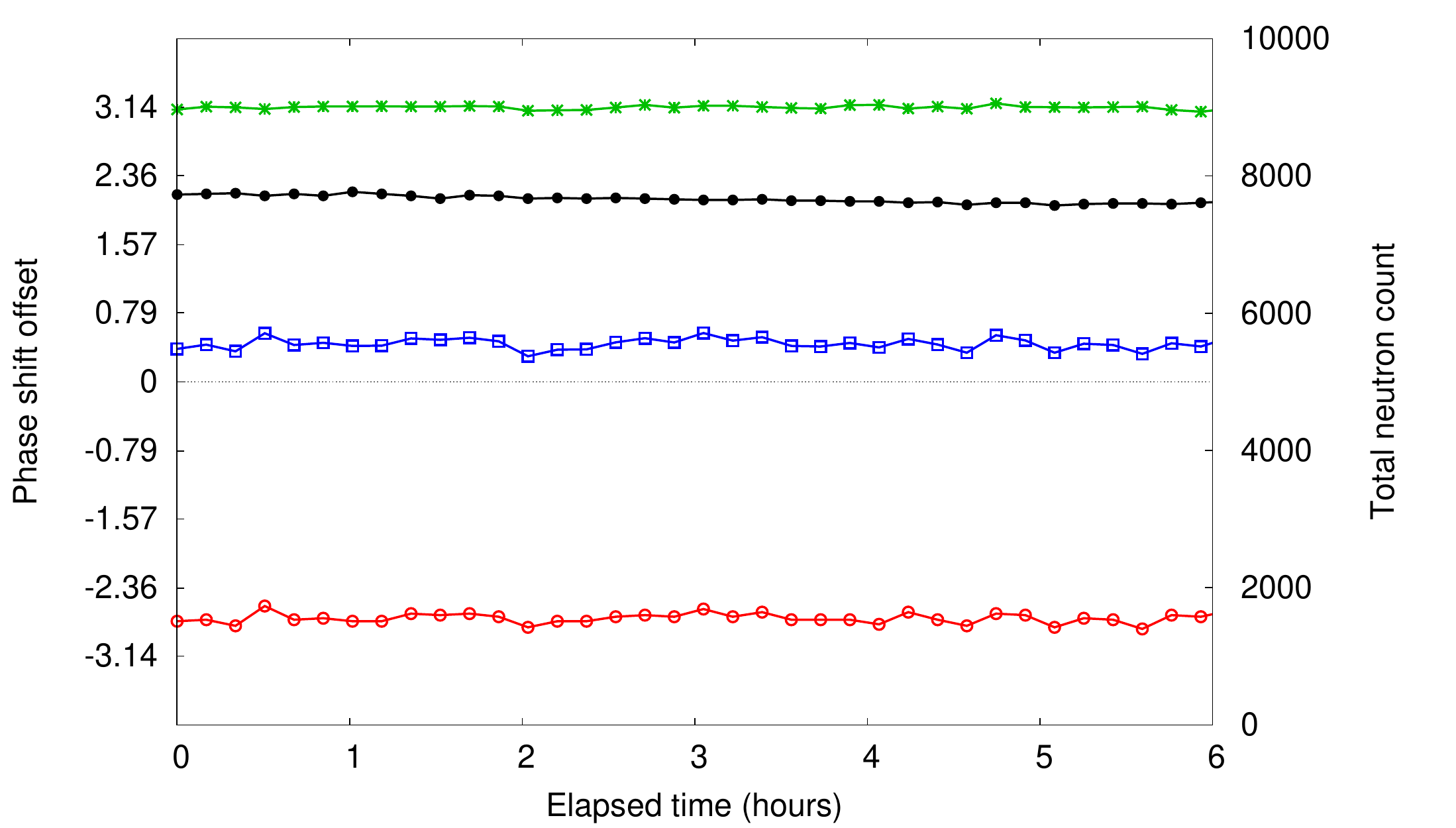} 
\caption{(color online) %
The variation of the phase shift offsets $\chi_\mathrm{O}$ ($\bigcirc$), $\chi_\mathrm{H}$ ($\Box$),
$\chi_\mathrm{H}-\chi_\mathrm{O}$ ($\ast$) and the total number of detected neutrons ($\bullet$)
as a function of real time. 
Lines through the data points are guides to the eye.
}
\label{fig3}
\end{center}
\end{figure}

A detailed quantum theoretical treatment of the interferometer depicted in Fig.~\ref{fig1}
is given in Refs.~\cite{RAUC74b,RAUC15}.
Assuming that the incident wave satisfies the Bragg condition for scattering by the
first crystal plate (BS0), the Laue-type interferometer acts as a two-path
interferometer~\cite{HORN86}.
The simplest description of the crystal plates BS0, BS1, BS2, and BS3
assumes that these plates act as beam splitters for plane waves of neutrons.
Then, the unitary matrix that relates the amplitudes of incoming, reflected and transmitted waves is given by
$U=\left(\begin{array}{cc}
        \phantom{-}\T^\ast &\R \\
        -\R^\ast & \T
\end{array}\right)$
where $\T$ and $\R$ denote the transmission and reflection coefficient, respectively,
and the probabilities to observe a particle leaving the interferometer
in the O- and H-beam are given by
\begin{eqnarray}
p_\mathrm{O}&=&2R^2T\left( 1+ \cos\chi \right)
,
\label{app2z}
\\ 
p_\mathrm{H}&=&R(T^2+R^2)\left(1 -\frac{2 RT}{T^2+R^2}\cos\chi \right)
,
\label{app2}
\end{eqnarray}
where $\chi$ is the relative phase shift between the two different paths that the neutrons can take, $R=|\R|^2$ and $T=|\T|^2=1-R$.
Note that $p_\mathrm{O}$ and $p_\mathrm{H}$ do not depend on the imaginary part of $\T$ or $\R$,
leaving only one free model parameter (e.g. $R$).
In the case of a 50-50 beam splitter ($T=R=1/2$), Eqs.~(\ref{app2z}) and (\ref{app2}) reduce to the familiar expressions
$p_\mathrm{O}=(1/2)\cos^2\chi/2$ and $p_\mathrm{H}=(1/2)\sin^2\chi/2$, respectively.
The extra factor one-half is due to the fact that if $T=R=1/2$,
one half of all incoming neutrons, that is the neutrons that are
transmitted by BS1 or BS2 (see Fig.~\ref{fig1}), leave the interferometer without being counted.

The expression in Eq.~(\ref{app2z}) shows that the normalized (to the maximum value) O-beam intensity
does not depend on the value of the reflectivity $R$.
Furthermore, Eq.~(\ref{app2z}) implies that the visibility of the O-beam is given by
\begin{eqnarray}
V_\mathrm{O}&\equiv&
\frac{\max_{\chi} p_\mathrm{O} - \min_{\chi} p_\mathrm{O}}{\max_{\chi} p_\mathrm{O} + \min_{\chi} p_\mathrm{O}} =1
.
\label{app2a}
\end{eqnarray}

Figure~\ref{fig2a} clearly shows that the visibility of the O-beam interference signal is not equal to one.
Therefore, instead of using Eqs.~(\ref{app2z}) and (\ref{app2}),
we take as a model for the neutron counts~\cite{RAUC15}
\begin{eqnarray}
N_\mathrm{O}&=&A_\mathrm{O}\left[ 1+ B_\mathrm{O}\cos(\Omega_\mathrm{O}X+\chi_\mathrm{O})  \right]
,
\label{app3z}
\\ 
N_\mathrm{H}&=&A_\mathrm{H}\left[1 + B_\mathrm{H}\cos(\Omega_\mathrm{H}X+\chi_\mathrm{H}) \right]
.
\label{app3}
\end{eqnarray}
Fitting Eqs.~(\ref{app3z}) and (\ref{app3}) to the data shown in Fig.~\ref{fig2a} yields
$A_\mathrm{O}\approx 2781$,
$A_\mathrm{H}\approx 4951$,
$B_\mathrm{O}\approx 0.74$,
$B_\mathrm{H}\approx 0.42$,
$\Omega_\mathrm{O}\approx0.60$,
$\Omega_\mathrm{H}\approx0.60$,
$\chi_\mathrm{O}\approx-2.75$,
and
$\chi_\mathrm{H}\approx0.37$.

For the statistical analysis of the time series to be presented later, it is
important to know over which time span the counts and phases are fairly constant
{\color{black}
as phase drifts due to thermal fluctuations are not unusual in such experiments~\cite{GEPP14}.
}
In Fig.~\ref{fig3}, we plot the phase shift offsets $\chi_\mathrm{O}$ ($\Box$), $\chi_\mathrm{H}$ ($\bigcirc$),
$\chi_\mathrm{H} - \chi_\mathrm{O}$ ($\ast$)
and the total number of detected neutrons as a function of the elapsed time. 
Note that $\chi_\mathrm{H} - \chi_\mathrm{O}\approx\pi$ implies that the maxima
in the H-counts are shifted by $180^\circ$ relative to those of the O-counts, in agreement with the quantum theoretical
prediction given by Eqs.~(\ref{app2z}) and (\ref{app2}).
During the first 6 hours of collecting data, the counts and phases are fairly constant.

\begin{figure}[t]
\begin{center}
\includegraphics[width=0.80\hsize]{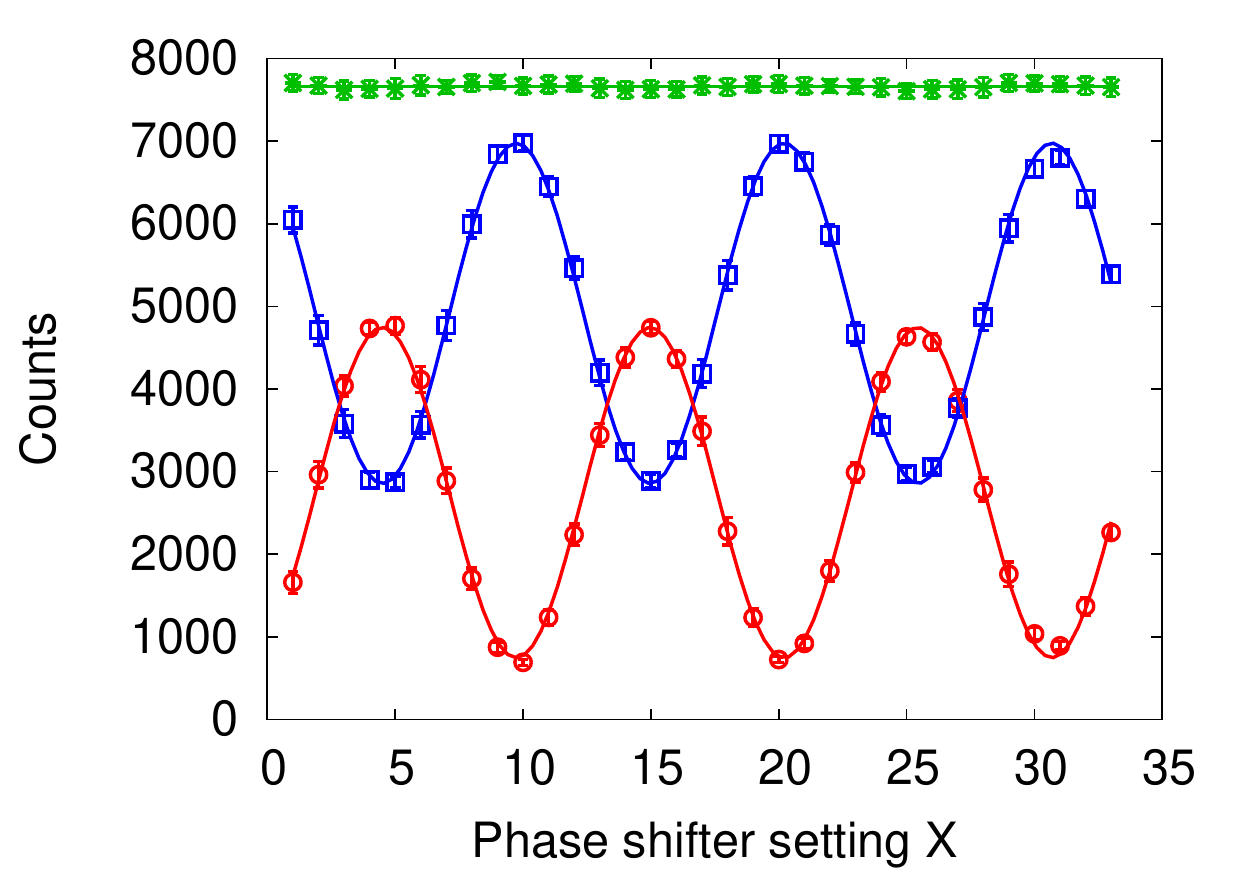} 
\caption{(color online) %
Neutron counts $N_{\mathrm{O}}$ and $N_{\mathrm{H}}$
as a function of the setting $X$, averaged over the 37 pairs of data files.
Open circles ($\bigcirc$): O-beam;
open squares ($\Box$): H-beam;
asterisks ($\ast$): sum of O- and H-beam counts.
Lines through the data points are obtained by fitting to the data as described in the text.
The standard deviations, indicated by the error bars, are of the size of the markers.
}
\label{fig2}
\end{center}
\end{figure}

Fitting the parameters that appear in Eqs.~(\ref{app3z}) and (\ref{app3}) to the
counts averaged over all experiments, 
we obtain
%
$A_\mathrm{O}\approx 2780$,
$A_\mathrm{H}\approx 4950$,
$B_\mathrm{O}\approx 0.74$,
$B_\mathrm{H}\approx 0.42$,
$\Omega_\mathrm{O}\approx0.60$,
$\Omega_\mathrm{H}\approx0.60$,
$\chi_\mathrm{O}\approx-2.71$,
and
$\chi_\mathrm{H}\approx0.43$, yielding $\chi\approx0.60 X$.
In Fig.~\ref{fig2} we present the data of counts averaged over all experiments
together with the results of the fitting procedure.

From Eqs.~(\ref{app2z}) and (\ref{app2}) and the values of the mean (over all settings) of the O- and
H-beam counts $A_\mathrm{O}$ and $A_\mathrm{H}$, we can estimate the reflectivity $R$ by solving
the quadratic equation
\begin{eqnarray}
\alpha=\frac{A_\mathrm{H}}{A_\mathrm{O}}=
\frac{\int_0^{2\pi} d\chi\; R(T^2+R^2)\left[1 - ({2 RT}/{(T^2+R^2)})\cos\chi \right]}{\int_0^{2\pi} d\chi\; 2R^2T\left[ 1+ \cos\chi \right]}
=\frac{(1-R)^2+R^2}{2R(1-R)}
,
\label{app4}
\end{eqnarray}
and we find that $\alpha\approx1.78$ and $R=(1+\sqrt{(\alpha-1)/(\alpha+1)})/2\approx0.76$ (or $0.24$ if we take the solution with the minus sign).
According to the quantum theoretical model, the relative amplitude of the H-beam fringes
should then be ${2 RT}/{(T^2+R^2)}=1/\alpha \approx 0.56$, which is not in good quantitative
agreement with the amplitude $B_\mathrm{H}\approx 0.42$ extracted from the experimental data.
In view of the  quantum model of crystal plates BS0, ..., BS3 that we used to estimate
$R$, this quantitative disagreement only calls for a more comprehensive
model of the Bragg scattering at the crystal plates~\cite{RAUC15}.
At any rate, Eqs.~(\ref{app3z}) and (\ref{app3}) with the values of the parameters
given above describe the experimental data very well.
The relative frequencies to detect a neutron in the O- or H-beam are, to a very good approximation, given by
\begin{eqnarray}
P_\mathrm{O}&=&\frac{A_\mathrm{O}}{A_\mathrm{O}+A_\mathrm{H}}
\bigg[ 1+ B_\mathrm{O}\cos(\Omega_\mathrm{O}X+\chi_\mathrm{O})  \bigg]
,
\label{app5z}
\\ 
P_\mathrm{H}&=&\frac{A_\mathrm{H}}{A_\mathrm{O}+A_\mathrm{H}}
\left[1 - \frac{B_\mathrm{O}A_\mathrm{O}}{A_\mathrm{H}}\cos(\Omega_\mathrm{O}X+\chi_\mathrm{O}) \right]
,
\label{app5}
\end{eqnarray}
where we used the empirical finding that $A_\mathrm{O}B_\mathrm{O}\approx A_\mathrm{H}B_\mathrm{H}$
and that $\chi_\mathrm{H}\approx\chi_\mathrm{O}+\pi$.

\subsection{Fitting the standard deviation}\label{section3c}

\begin{figure}[t]
\begin{center}
\includegraphics[width=0.45\hsize]{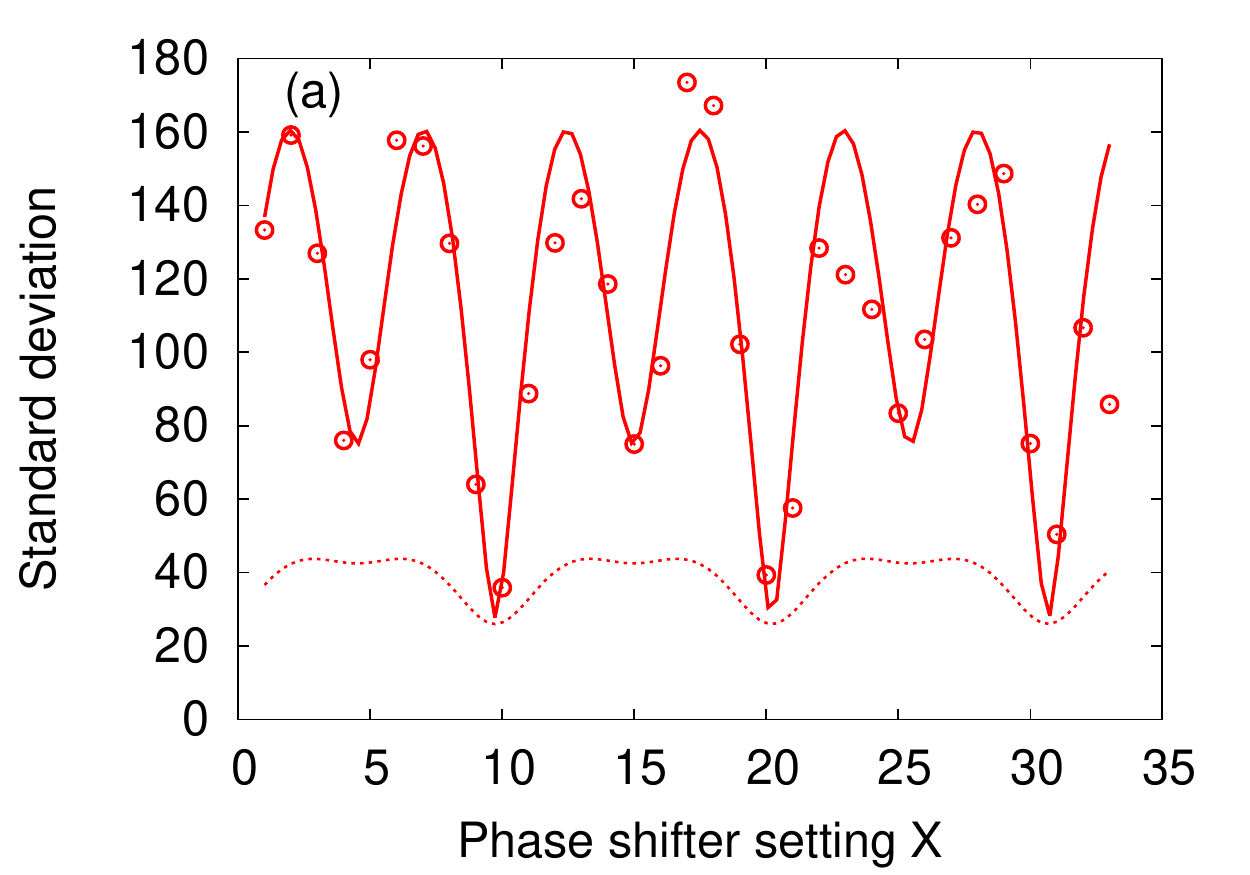} 
\includegraphics[width=0.45\hsize]{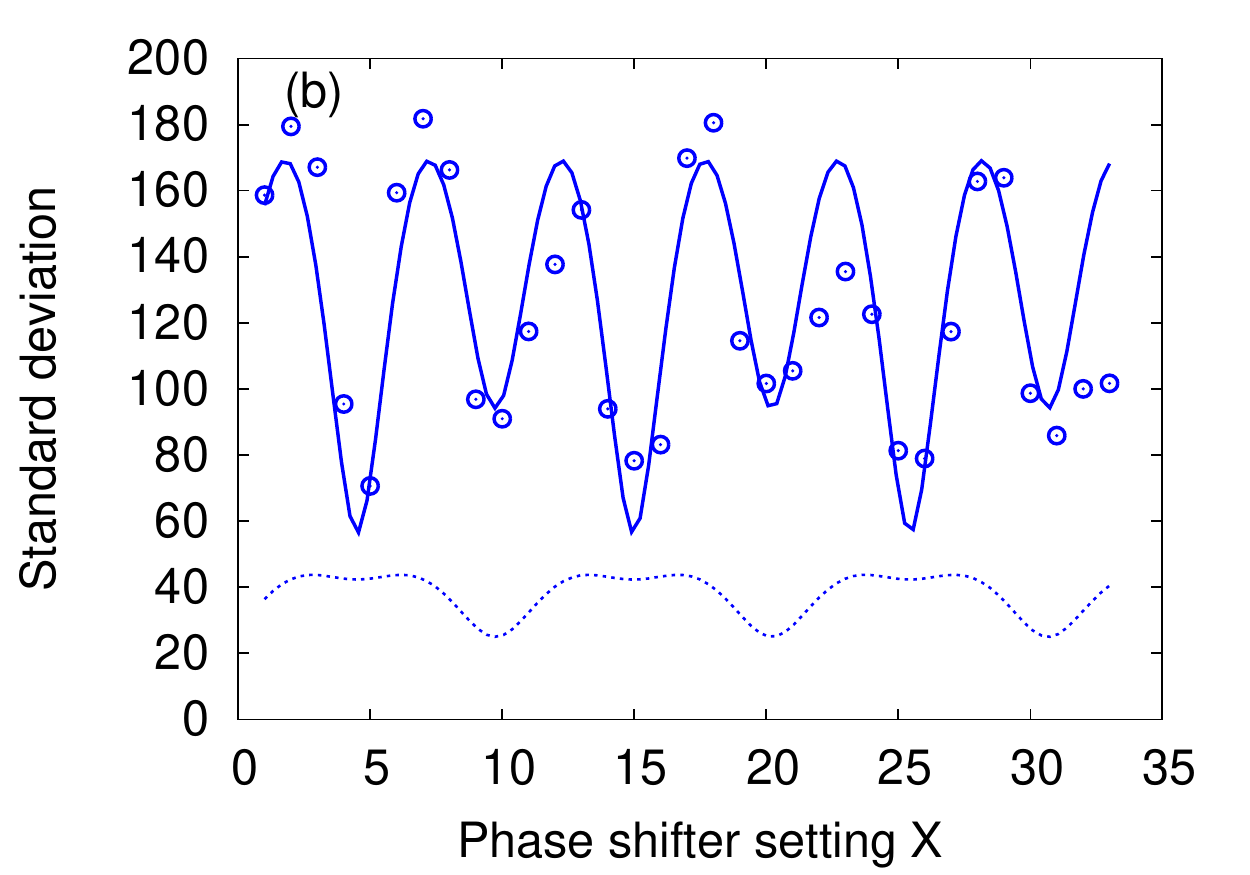}
\caption{(color online) %
{\color{black}
Standard deviation of the neutron counts as a function of the setting $X$,
as obtained from the 37 pairs of data files.
(a) $\bigcirc$: experimental O-beam data.
Solid line: $(\langle N^2_{\mathrm{O}} \rangle-\langle N_{\mathrm{O}}\rangle^2)^{1/2}$
as obtained from the theoretical model Eqs.~(\ref{app6c}) and~(\ref{app6d}) accounting
for fluctuations of the phase shift.
Dashed line: $(\langle N^2_{\mathrm{O}} \rangle-\langle N_{\mathrm{O}}\rangle^2)^{1/2}
=\left(NP_\mathrm{O}P_\mathrm{H}\right)^{1/2}$
as obtained from the standard model Eq.~(\ref{app6x}), using the value of
$P_\mathrm{O}$ obtained from the experimental O-beam counts.
(b) Same as (a) but for the H-beam instead of the O-beam data.
}
}
\label{fig4}
\end{center}
\end{figure}

In Fig.~\ref{fig2}, the error bars are hardly discernible but as shown in Fig.~\ref{fig4},
the standard deviations of the counts as a function of $X$ explicitly display
a nontrivial periodic dependence on $X$.
The standard deviation is at a minimum for those settings at which the corresponding counts reach a maximum or a minimum.

We make contact with a probabilistic model, {\color{black} in this case} quantum theory, by
assigning the values of the relative frequencies given by Eqs.~(\ref{app5z}) and (\ref{app5})
to the probabilities of the corresponding events.
{\color{black}
Quantum theory postulates that the detected events (represented by $x=+1(-1)$
for an O (H) detection event, respectively) are statistically independent random variables.
Therefore, for each of the 37 repetitions of the experiment,
we expect that
the probability to observe $N_\mathrm{O}$ counts is given by the binomial distribution~\cite{BARL89}
\begin{eqnarray}
p(N_\mathrm{O})&=&\frac{N!}{N_\mathrm{O}!(N-N_\mathrm{O})!}P_\mathrm{O}^{N_\mathrm{O}}(1-P_\mathrm{O})^{N-N_\mathrm{O}}
,
\label{app6}
\end{eqnarray}
where $N=A_\mathrm{O}+A_\mathrm{H}$ is the total number of events observed in one run.
From Eq.~(\ref{app6}), it follows that for the same run, the averages of $N_\mathrm{O}$ and $N^2_\mathrm{O}$ are given by
\begin{eqnarray}
\langle N_\mathrm{O}\rangle =N P_\mathrm{O}
\quad,\quad
\langle N^2_\mathrm{O}\rangle =N^2 P^2_\mathrm{O}+ NP_\mathrm{O}(1-P_\mathrm{O})
,
\label{app6x}
\end{eqnarray}
respectively.

Plotting the standard deviations
$(\langle N^2_\mathrm{O}\rangle-\langle N_\mathrm{O}\rangle^2)^{1/2}=$
$(\langle N^2_\mathrm{H}\rangle-\langle N_\mathrm{H}\rangle^2)^{1/2}=(NP_\mathrm{O}P_\mathrm{H})^{1/2}$
obtained by using the values of the parameters
determined by fitting to the experimentally observed counts
yields curves that are very different from those of the standard deviations computed from the
data obtained by the 37 repetitions of the experiment (see below).
Therefore, the observed data cannot be described by a simple Bernoulli process.
Of course, this process does not capture all the physical influences that affect the O- and H-beam counts, such as
fluctuations in the total neutron count $N=A_\mathrm{O}+A_\mathrm{H}$ from run to run
(see the asterisks in Fig.~\ref{fig2})
and fluctuations $\epsilon$ of the phase shift at a fixed setting $X$.
{\color{black}
The internal phase of the interferometer varies by about 0.0349 
due to reflection effects (Pendell\"osung structure) of the interferometer crystal~\cite{RAUC15}.
}

We incorporate the effect of the two types of fluctuations into the probabilistic model
by assuming that they are described by the probability distribution $p(N)$ and $p(\epsilon)$, respectively,
and that the individual O- and H-detection events are described by Bernoulli processes with success probabilities
}
\begin{eqnarray}
\widehat P_\mathrm{O}(X,\epsilon)&=&\frac{A_\mathrm{O}}{A_\mathrm{O}+A_\mathrm{H}}
\bigg[ 1+ B_\mathrm{O}\cos(\Omega_\mathrm{O}X+\chi_\mathrm{O}+\epsilon)  \bigg]
,
\label{app6y}
\\
\widehat P_\mathrm{H}(X,\epsilon)&=&\frac{A_\mathrm{H}}{A_\mathrm{O}+A_\mathrm{H}}
\left[1 - \frac{B_\mathrm{O}A_\mathrm{O}}{A_\mathrm{H}}\cos(\Omega_\mathrm{O}X+\chi_\mathrm{O}+\epsilon) \right]
.
\label{app6z}
\end{eqnarray}

Temporarily writing $\widehat P=\widehat P(\epsilon)=\widehat P_{\mathrm{O}}(X,\epsilon)$,
we have for the average O-count
\begin{eqnarray}
\langle N_{\mathrm{O}} \rangle&=& \sum_{N}
\int N \widehat P(\epsilon) p(N) p(\epsilon) \;d\epsilon  = \langle N\rangle_N \langle \widehat P\rangle_\epsilon
,
\label{app6a}
\end{eqnarray}
where $\langle .\rangle_N$ and $\langle .\rangle_\epsilon$ denote the expectation value with respect to the
probability distribution $p(N)$ and $p(\epsilon)$, respectively.
For the average of the O-counts squared we have
\begin{eqnarray}
\langle N^2_{\mathrm{O}} \rangle&=& \sum_{N} \int  \left[ N(N-1) \widehat P^2(\epsilon) + N \widehat P(\epsilon)\right]
 p(\epsilon) p(N) \;d\epsilon
=\langle N(N-1)\rangle_N \langle \widehat P^2\rangle_\epsilon + \langle N\rangle_N \langle \widehat P\rangle_\epsilon
\nonumber \\
&=& (\langle N^2\rangle_N^{\phantom{2}}  - \langle N\rangle_N^2 )(\langle \widehat  P^2 \rangle_\epsilon^{\phantom{2}}
-\langle \widehat P \rangle_\epsilon^2)
+ \langle N\rangle_N^{\phantom{2}} ( \langle N \rangle_N^{\phantom{2}}-1)
(\langle \widehat P^2 \rangle_\epsilon^{\phantom{2}} -\langle \widehat P \rangle_\epsilon^2)
\nonumber \\
&&+ \langle \widehat P\rangle_\epsilon^2 (\langle N^2 \rangle_N^{\phantom{2}} -\langle N \rangle_N^2)
+ \langle N\rangle_N^{\phantom{2}} \langle \widehat P \rangle_\epsilon (1-\langle \widehat P \rangle_\epsilon^{\phantom{2}})
+ \langle N\rangle_N^2 \langle \widehat P \rangle_\epsilon^2
,
\label{app6b}
\end{eqnarray}
such that the variance on the O-counts is given by
\begin{eqnarray}
\langle N^2_{\mathrm{O}} \rangle-\langle N_{\mathrm{O}}^{\phantom{2}} \rangle^2
&=& \sigma_N^2 \sigma_{\widehat P}^2 + \langle \widehat P \rangle_\epsilon^2\sigma_N^2
+ \langle N \rangle_N^{\phantom{2}}(\langle N \rangle_N^{\phantom{2}}-1) \sigma_{\widehat P}^2
+ \langle N \rangle_N\langle \widehat P \rangle_\epsilon (1-\langle \widehat P \rangle_\epsilon)
,
\label{app6c}
\end{eqnarray}
where $\sigma_N^2 = \langle N^2\rangle_N^{\phantom{2}}  - \langle N\rangle_N^2$ and
$\sigma_{\widehat P}^2=\langle \widehat P^2 \rangle_\epsilon^{\phantom{2}} -\langle \widehat P \rangle_\epsilon^2$.
The variance on the H-count is given by Eq.~(\ref{app6c})
with the subscript O replaced by H and $\widehat P$ replaced by
$\widehat P= \widehat P(\epsilon)=\widehat P_{\mathrm{H}}(X,\epsilon)$.

{\color{black}
Assuming that there are no fluctuations in the phase ($\sigma_{\widehat P}^2=0$)
and that the total number of counts varies from run to run in a way described by
a Poisson process ($\sigma_N^2=\langle N \rangle_N$),
Eq.~(\ref{app6c}) reduces to
$\langle N^2_{\mathrm{O}} \rangle-\langle N_{\mathrm{O}}^{\phantom{2}} \rangle^2
=\langle N \rangle_N P_{\mathrm{O}}$, meaning that
the standard deviation on the counts is, up to a factor,
given by the square root of the counts~\cite{RAUC15}, in blatant contradiction with the data analyzed in this paper.
It other words, the assumption that there are no fluctuations in the phase may not be correct.
As a matter of fact, the estimated phase fluctuates from run to run, see Fig.~\ref{fig3} and
also Fig.~2.10 in Ref.~\cite{RAUC15}.
Whether or not these long-term phase fluctuations have a noticeable effect on the standard deviation
depends on the relative weight of the $\sigma_{\widehat P}^2$ terms in Eq.~(\ref{app6c}).
Apparently, for the data analyzed in this paper, the contribution of the $\sigma_{\widehat P}^2$
terms to the standard deviation is significant, see Fig.~\ref{fig4}.
}

As a simple model for the random fluctuations in the phase shift for a
fixed setting $X$, we take a uniform distribution over a small interval $[-\epsilon_0,\epsilon_0]$.
Then, we find that
\begin{eqnarray}
\langle \widehat P_\mathrm{O} \rangle_\epsilon&=&\frac{A_\mathrm{O}}{A_\mathrm{O}+A_\mathrm{H}}
\bigg[ 1+ (1-\frac{\epsilon_0^2}{6})B_\mathrm{O}\cos(\Omega_\mathrm{O}X+\chi_\mathrm{O})  \bigg] +{\cal O}(\epsilon_0^3)
,
\nonumber \\
\langle \widehat P_\mathrm{H} \rangle_\epsilon&=&\frac{A_\mathrm{H}}{A_\mathrm{O}+A_\mathrm{H}}
\left[1 - (1-\frac{\epsilon_0^2}{6})\frac{B_\mathrm{O}A_\mathrm{O}}{A_\mathrm{H}}
\cos(\Omega_\mathrm{O}X+\chi_\mathrm{O}) \right] +{\cal O}(\epsilon_0^3)
,
\nonumber \\
\sigma_{\widehat P}^2&=&\frac{\epsilon_0^2}{3}\left(
\frac{A_{\mathrm{O}}B_{\mathrm{O}}}{A_{\mathrm{O}}+B_{\mathrm{O}}}\right)
\sin^2 (\Omega_\mathrm{O}X+\chi_\mathrm{O})+{\cal O}(\epsilon_0^3)
,
\label{app6d}
\end{eqnarray}
where the latter holds for both the O- and H-counts.

{\color{black}
The numerical results for
$(\langle N^2_{\mathrm{O}} \rangle-\langle N_{\mathrm{O}}^{\phantom{2}} \rangle^2)^{1/2}$
are shown in Fig.~\ref{fig4} as the solid line.
The standard deviation of the neutron counts in the H-beam and
$(\langle N^2_{\mathrm{H}} \rangle-\langle N_{\mathrm{H}}^{\phantom{2}} \rangle^2)^{1/2}$
show similar agreement (not shown).}
The values of the parameters used are
$A_\mathrm{O}=2745$,
$A_\mathrm{H}=4916$,
$B_\mathrm{O}=0.73$,
$B_\mathrm{H}=0.42$,
$\Omega_\mathrm{O}=0.60$,
$\chi_0=-2.7$,
and
$\epsilon_0=0.13$.
This is a slightly different set of parameters than the one used to draw the lines in
Fig.~\ref{fig2} but as the changes in the new plot of the counts are not visible, we do not show it.

\begin{figure}[t]
\begin{center}
\includegraphics[width=0.90\hsize]{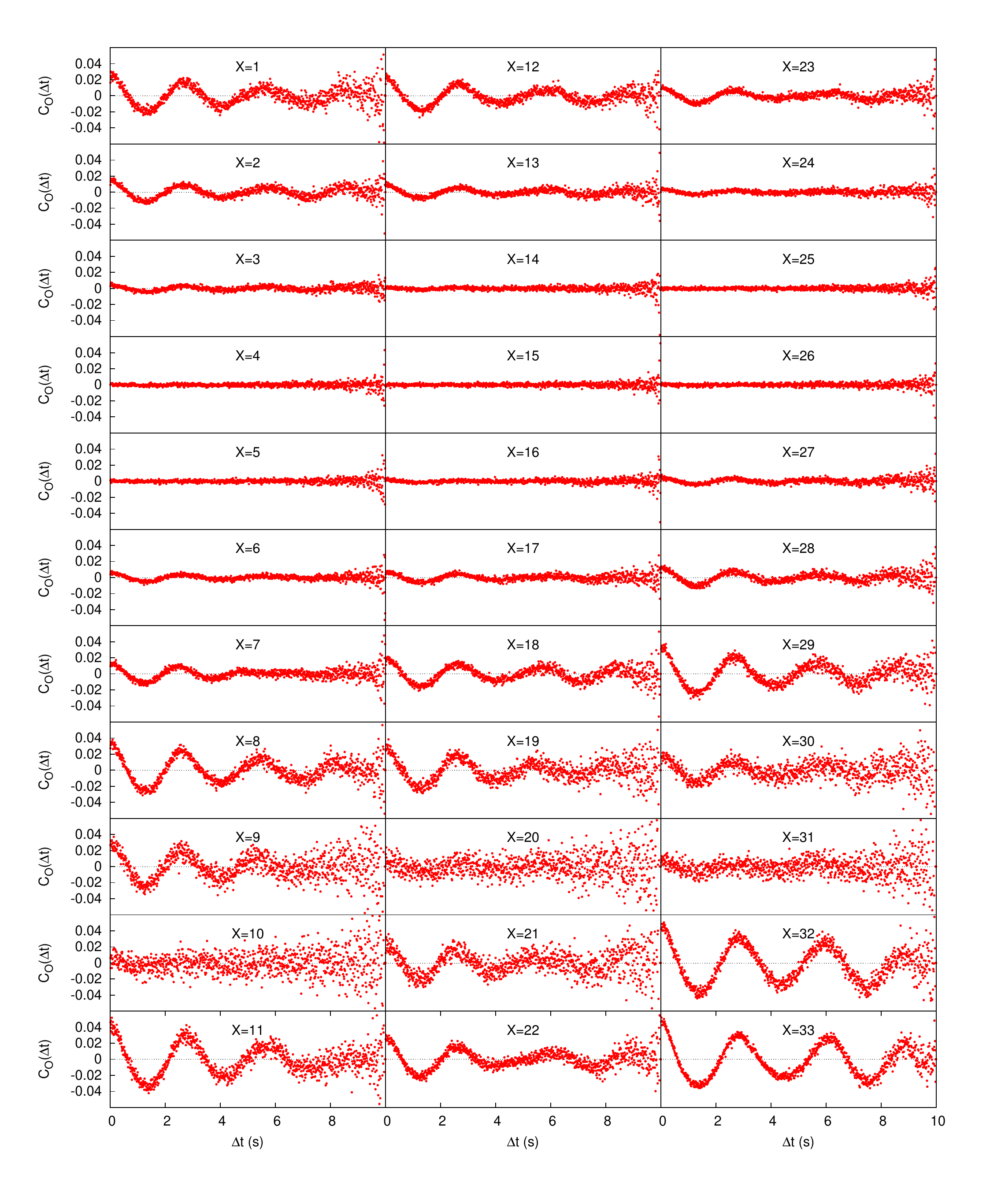} 
\caption{(color online) %
The correlation $C_{\mathrm{O}}(\Delta t)$ of the time-differences between the O-beam detection events for
different phase shifter settings $X$ as a function of the time difference $\Delta t$,
computed from the 37 pairs of data files.
{\color{black}Fitting the function $f(\Delta t)=a\exp(-b \Delta t)\cos(2\pi\Delta t/{\cal T})$
to the data yields ${\cal T}\approx2.8 \;\mathrm{s}$ for the period of the oscillations in the correlation
$C_{\mathrm{O}}(\Delta t)$.}
}
\label{fig5}
\end{center}
\end{figure}

\begin{figure}[t]
\begin{center}
\includegraphics[width=0.90\hsize]{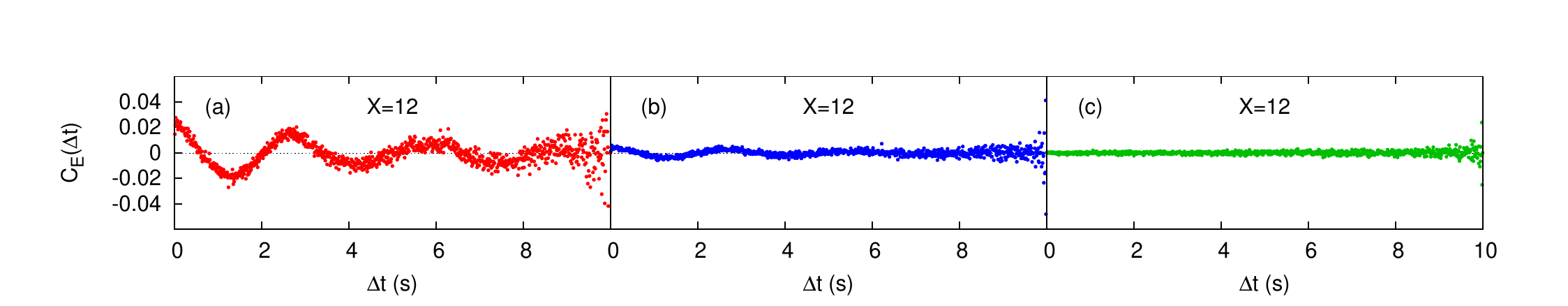} 
\caption{(color online) %
The correlation $C_{\mathrm{E}}(\Delta t)$ of the time-differences
of the O-beam (a, E=O), H-beam (b, E=H) and O-beam or H-beam (c, E=OH) detection events
for the same phase shifter setting $X=12$ as a function of the time difference $\Delta t$,
computed from the 37 pairs of data files.
}
\label{fig5y}
\end{center}
\end{figure}

\begin{figure}[b]
\begin{center}
\includegraphics[width=0.80\hsize]{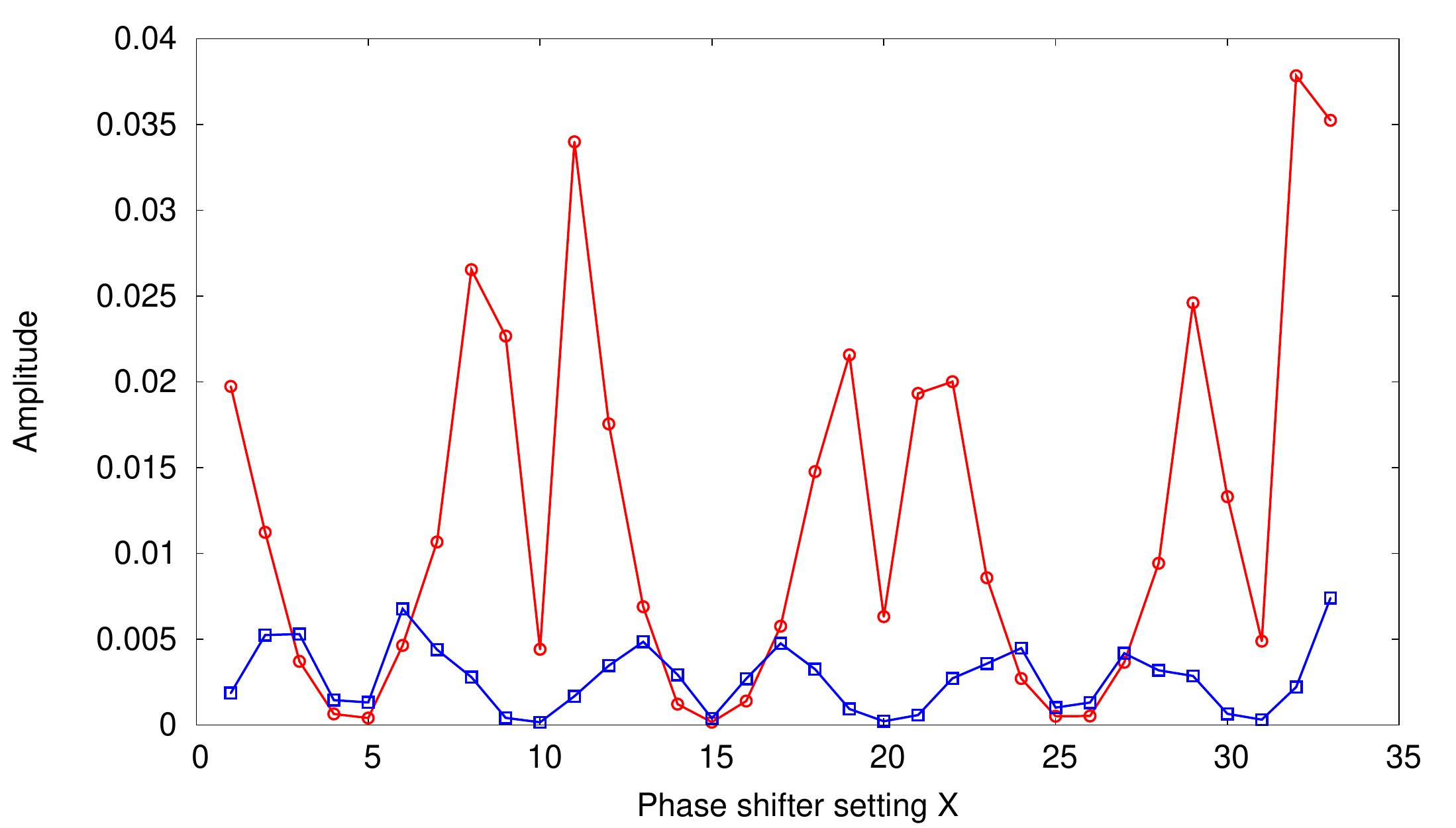} 
\caption{(color online) %
The maximum amplitudes of the oscillations in
$C_{\mathrm{O}}(\Delta t)$ ($\bigcirc$)
and
$C_{\mathrm{H}}(\Delta t)$ ($\Box$),
computed from the experimental data,
as a function of the phase shifter setting $X$.
Lines through the data points are guides to the eye.
}
\label{fig5z}
\end{center}
\end{figure}

\section{Analysis of the time series}\label{section4}
\subsection{Correlation of the time differences}\label{section4a}

{\color{black}
We denote the $n$th moment of the time difference of
two successive neutron detection events in either the O- or H-beam by
\begin{eqnarray}
\langle (\Delta t)^n\rangle &\equiv& \frac{1}{M}\sum_{i=1}^{M} \left(\Delta t_{i}^{\mathrm{(E)}}\right)^n
,
\label{app7a}
\end{eqnarray}
where we introduce the subscript $\mathrm{E}\in\{\mathrm{O},\mathrm{H},\mathrm{OH}\}$ to indicate
that we consider only the O-, H- or (O or H)-beam detection events, respectively.
From the time-stamp data, we find for the average time $\langle \Delta t \rangle\approx1.3\;\mathrm{ms}$,
which is much larger than
the time it takes for a neutron to traverse the interferometer (which is of the order of $0.1\mathrm{ms}$~\cite{RAUC15}).
Therefore, it is very reasonable to assume that at any time, there is at most one neutron in the region of
space occupied by the interferometer~\cite{RAUC15}.
In other words, if the experiment yields an interference signal, this signal is built up
by single neutrons that travel though the interferometer one-by-one~\cite{RAUC74a,RAUC15}.
{\color{black}If the distribution of the time differences is exponential (as in the case of a Poisson process),
the averages of the first three moments of the time differences are related by
$\langle \Delta t\rangle =(\langle (\Delta t)^2\rangle/2)^{1/2} =(\langle (\Delta t)^3\rangle/6)^{1/3}$.
Using the experimental data, we find that within 1\%,
$\langle \Delta t\rangle \approx (\langle (\Delta t)^2\rangle/2)^{1/2} \approx (\langle (\Delta t)^3\rangle/6)^{1/3}$,
suggesting that $\Delta t_i^{(\mathrm{OH})}$ may indeed be described by an exponential distribution.
}
Plotting (data not shown) the distribution of the frequencies with which the $\Delta t_i^{(\mathrm{E})}$'s
(for $\mathrm{E}\in\{\mathrm{O},\mathrm{H},\mathrm{OH}\}$)
occur indeed shows the exponential dependence,
in concert with the finding that the detection times can be described by a Poisson process~\cite{RAUC15}.

A key feature of quantum theory and also of a Poisson process is that the events are independent~\cite{GRIM01}.
We can test the hypothesis that the detection events are independent by
studying the time series of the $ \Delta t_i$'s in more detail.

For each of the $n$ time-stamp files and for each setting $X$, we calculate the averages
\begin{eqnarray}
\langle \Delta t^{(\mathrm{E},k)} \rangle
&\equiv&
\frac{1}{M^{(\mathrm{E})}-k}\sum_{i=1}^{M^{(\mathrm{E})}-k} \Delta t^{(\mathrm{E})}_{i+k}\;,
\nonumber \\
\langle \left(\Delta t^{(\mathrm{E},k)}\right)^2 \rangle
&\equiv& \frac{1}{M^{(\mathrm{E})}-k}\sum_{i=1}^{M^{(\mathrm{E})}-k} \left(\Delta t^{(\mathrm{E})}_{i+k}\right)^2\;,
\nonumber \\
\langle \Delta t^{(\mathrm{E},0)} \Delta t^{(\mathrm{E},k)} \rangle
&\equiv&
\frac{1}{M^{(\mathrm{E})}-k}\sum_{i=1}^{M^{(\mathrm{E})}-k}
\Delta t^{(\mathrm{E})}_{i}\Delta t^{(\mathrm{E})}_{i+k}
\;,
\label{app7}
\end{eqnarray}
where $M^{(\mathrm{E})}$ is the number of time differences per setting $X$
and $k=0,1,\ldots,M_{\mathrm{E}}-1$.
We use these averages to compute the correlation~\cite{BARL89}
\begin{eqnarray}
C_{\mathrm{E}}(\Delta t) =
C_{\mathrm{E}}(k\langle \Delta t \rangle) \equiv
\frac{\langle \Delta t^{(\mathrm{E},0)} \Delta t^{(\mathrm{E},k)} \rangle
- \langle \Delta t^{(\mathrm{E},0)}\rangle \langle \Delta t^{(\mathrm{E},k)} \rangle}{
\sqrt{
[\langle \left(\Delta t^{(\mathrm{E},0)}\right)^2\rangle-\langle \Delta t^{(\mathrm{E},0)} \rangle^2]
[\langle \left(\Delta t^{(\mathrm{E},k)}\right)^2\rangle-\langle \Delta t^{(\mathrm{E},k)} \rangle^2]}}
,
\label{app8}
\end{eqnarray}
which is a dimensionless number between -1 and +1.
}
In order to relate the index $k$ to a real time difference, we have introduced the variable
$\Delta t= k\langle\Delta t\rangle$.
Recall that $0\le \Delta t \le 10\,\mathrm{s}$.
Applied to the data files,
the result of this whole operation is
a set of 37 values of $C_{\mathrm{E}}(\Delta t)$ for each $X=1,\ldots,33$ and
for the time-differences between the O-, H- and (O or H)-beam detection events.
Note that if the $\Delta t_{i}^{(\mathrm{E})}$'s were independent random variables,
we would have $C_{\mathrm{E}}(\Delta t)=0$ for all $\Delta t >0$.

In Fig.~\ref{fig5}, we present the 33 plots as obtained from the
time-differences between the O-beam detection events.
These pictures show that
\begin{enumerate}
\item
For several values of the phase shifter setting $X$, the correlation $C_{\mathrm{O}}(\Delta t)$ exhibits
oscillatory behavior. The period of the oscillations in the correlations is
{\color{black}${\cal T}\approx2.8\;\mathrm{s}$},
independent of $X$ with amplitudes depending on $X$.
\item
As $\Delta t\rightarrow 10\;\mathrm{s}$, the scattering of the points in the plots
increases due to the fact that as $k\rightarrow M$, there are only a few data points per bin ($0.01\;\mathrm{s}$ in this paper).
\end{enumerate}
The plots of $C_{\mathrm{H}}(\Delta t)$ display less pronounced
oscillations and those of $C_{\mathrm{OH}}(\Delta t)$ exhibit none at all,
as illustrated in Fig.~\ref{fig5y}.
Figure~\ref{fig5z} shows the maximum amplitudes of the oscillations in $C_{\mathrm{O}}(\Delta t)$ and
$C_{\mathrm{H}}(\Delta t)$ as a function of the phase shifter setting $X$.
As in the case of the standard deviations, there is a marked regularity in these data.

\subsection{Correlation in the time series of O- and H-counts}\label{section4b}

Additional evidence for the existence of long-time correlations in the detection events
comes from the relative frequency with which say an O(H)-beam detection event (represented by $x_i=-1(+1)$)
correlates with another O (H)-beam detection event (represented by $x_{i+k}=-1(+1)$) separated by $k-1$ other events
(see also Table~\ref{tab1}).
This kind of information is contained in the correlation
{\color{black}
\begin{eqnarray}
C_{\mathrm{x}}(\Delta t)&=& C(k\langle\Delta t\rangle)\equiv
\frac{\langle x^{(0)} x^{(k)} \rangle- \langle x^{(0)} \rangle \langle x^{(k)} \rangle}{
\sqrt{
[\langle (x^{(0)})^2 \rangle-\langle x^{(0)}\rangle^2]
[\langle (x^{(k)})^2 \rangle-\langle x^{(k)}\rangle^2]
}}
,
\label{app9}
\\
\noalign{\noindent where}
\langle x^{(k)} \rangle&\equiv& \frac{1}{M-k}\sum_{i=1}^{M-k} x_i
\quad,\quad
\langle (x^{(k)})^2 \rangle\equiv \frac{1}{M-k}\sum_{i=1}^{M-k} x_i^2
\quad,\quad
\langle x^{(0)} x^{(k)} \rangle\equiv \frac{1}{M-k}\sum_{i=1}^{M-k} x_i x_{i+k}
.
\end{eqnarray}
}
As before, if the $x_i$'s were independent random variables, we would have $C_{\mathrm{x}}(\Delta t)=0$ for all
$\Delta t >0$.

\begin{figure}[t]
\begin{center}
\includegraphics[width=0.90\hsize]{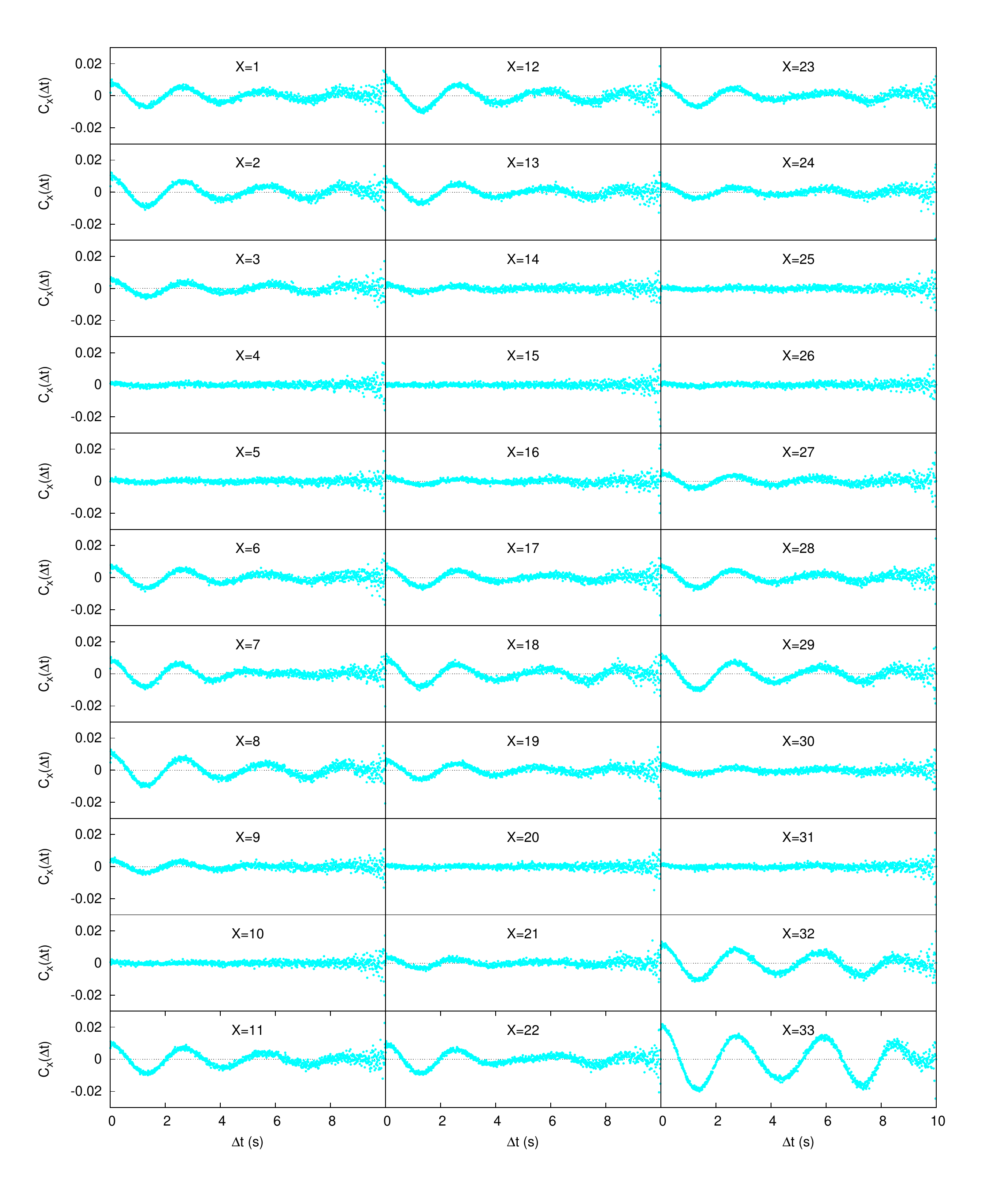} 
\caption{(color online) %
The correlation $C_{\mathrm{x}}(\Delta t)$ of
the variables $x_i=\pm1$ labeling O- or H-beam detection events (see Table~\ref{tab1})
for different phase shifter settings $X$ as a function of the time difference $\Delta t$,
computed from the data.
{\color{black}Fitting the data to the function $f(\Delta t)=a\exp(-b \Delta t)\cos(2\pi\Delta t/{\cal T})$
yields ${\cal T}\approx2.8 \;\mathrm{s}$ for the period of the oscillations in the correlation
$C_{\mathrm{x}}(\Delta t)$.}
}
\label{fig6}
\end{center}
\end{figure}

In Fig.~\ref{fig6}, we present the 33 plots obtained by computing Eq.~(\ref{app9})
using the same data files as those used to produce Fig.~\ref{fig5}.
The amplitudes of the oscillations in $C_{\mathrm{x}}(\Delta t)$
are somewhat less pronounced than those in $C_{\mathrm{O}}(\Delta t)$ for the corresponding setting $X$.
Qualitatively, the $x_i-x_{i+k}$ correlations show the same features as the correlations of the time differences,
suggesting that also the $x_i$'s cannot be modeled by independent random variables.

\clearpage

{\color{black}
\subsection{Discussion}

The correlations computed from the time-stamped data of repeated measurments of 10 s each exhibit oscillations, see Figs.~\ref{fig5} and~\ref{fig6}.
This observation is in conflict with a model based on independent random variables.

The same analysis was applied to the time series recorded on 19
June 2015 using polarized neutrons.
These data have been obtained by using 45 s (instead of 10 s for the
experiment with unpolarized neutrons) per setting $X$, resulting in about 1.5 million time stamps
collected over a time span of 3 hours and 22 minutes.
The data for the correlations $C_{\mathrm{O}}(\Delta t)$ and $C_{\mathrm{x}}(\Delta t)$ (see next section)
suggest that the same oscillations are also present in this case.
However, the statistics is too poor to allow for a reliable extraction of the period of the oscillations.
The experiment with polarized neutrons was repeated on 25 and 26 June 2015,
with 60 seconds instead of 45 s per setting $X$, collecting about 2.1 million neutrons in about
9 hours and 13 minutes. The statistics of these data is not as good as for the data taken on
19 June 2015 and the evidence for oscillating correlations is very weak.

Computing the correlations using only the first five (twenty) seconds or
the last five (twenty) seconds of data collected on 30 April 2015 (19 June 2015) per phase shifter setting
has no significant effect on the visibility of the oscillations and therefore
we do not show the corresponding figures.

The correlations obtained from the experiments with polarized neutrons show oscillations with much less visibility than the oscillations observed in the experiments with unpolarized neutrons.
The former experiments were carried out with 45 and 60 s measurement time per setting.
The latter experiments were carried out with 10 s measurement time per setting.
Although the statistics of the former does not allow us
to make a definite statement, the experimental data suggests that
in all cases, the correlations $C_{\mathrm{O}}(\Delta t)$ and $C_{\mathrm{x}}(\Delta t)$ show oscillations
with a period of about $2.8\,$s.

}

\section{Oscillating phase difference}\label{section5}

The observed oscillations in the correlations $C_{\mathrm{x}}(\Delta t)$
of the time series of the O- and H-counts as well as the ones observed
in the correlations $C_{\mathrm{E}}(\Delta t)$ of the time-differences exhibit features
that defy an explanation in terms of a quantum model of an ideal single-neutron interferometer.
Indeed, in the latter model, these correlations are zero.
In the case $C_{\mathrm{x}}(\Delta t)$ for which we can associate projection operators
to the observation of an O- or H-event we can, at least in principle, make use of the quantum formalism.
However, in the case of $C_{\mathrm{E}}(\Delta t)$ there is no such association with operators
of the quantum formalism~\cite{BALL03}.
Indeed, quantum theory treats time as a parameter and not as an observable represented
by an operator. The times at which the O- or H-detectors click refer
to the readings of a clock, not to the measurement of some special ``time variable''.
In this sense, a calculation of $C_{\mathrm{E}}(\Delta t)$ is out of the scope of quantum theory.

We now show that the ``unexplained'' features can readily be
reproduced by a single-neutron interferometer simulation model in which the phase difference oscillates in time.
In other words, instead of having $\Omega_\mathrm{O}X + \chi_\mathrm{O}+\epsilon$
as the expression of the phase difference in Eqs.~(\ref{app6y}) and ~(\ref{app6z}), we assume that we have

\begin{eqnarray}
\widetilde \Omega(t) = \Omega_\mathrm{O}X + \chi_\mathrm{O}+\epsilon+Y\sin(\omega (t-t_0)),
\label{sec4a0}
\end{eqnarray}
where $Y$ is regarded as a new adjustable parameter and
{\color{black}$\omega=2\pi/{\cal T}\approx2\pi/2.8\;\mathrm{s}^{-1}$}
is the angular frequency of the oscillations in the correlation of time-differences.
Note that we have taken as the period of these oscillations {\color{black}${\cal T}\approx 2.8 \;\mathrm{s}$},
the value observed in the experiment. The presence of the time offset $t_0$ accounts
for the lack of knowledge about the time
at which the oscillation of the phase shift sets in, relative to the arrival time of the first
neutron for the setting $X$.
{\color{black} As in our model for the standard deviation (see section~\ref{section3c}),
$\epsilon$ is chosen uniformly at random from the range $[-\epsilon_0,\epsilon_0]$ for each event.
}
Simulations (data not shown) with $\omega t_0$ chosen randomly do not reveal any significant dependence on $\omega t_0$.
Therefore, for all practical purposes we may think of $t_0$ being zero.

As the phase given by Eq.~(\ref{sec4a0}) explicitly depends on time,
the time at which a neutron impinges on the first beam splitter BS0 of the interferometer
affects its frequency to be detected in either the O-beam or H-beam detector.
Therefore, in contrast to the standard treatment in terms of a stationary description,
only a description that is intrinsically time-dependent can reproduce the oscillations in
the correlation of the time-differences $C_{\mathrm{O}}(\Delta t)$.
In essence, this enforces a description in terms of discrete events.

We consider two, conceptually very different, simulation models.
Common to both models is that the times of the incident neutrons are generated
according to a Poisson process, such that the average number of detected neutrons in the simulation is close
to the one observed in experiment.
The specifications of the models for
the propagation of the individual neutrons through the interferometer and their detection process
is given in the following two subsections.

\subsection{Wave function collapse model}\label{section5a}

\begin{figure}[t]
\begin{center}
\includegraphics[width=0.90\hsize]{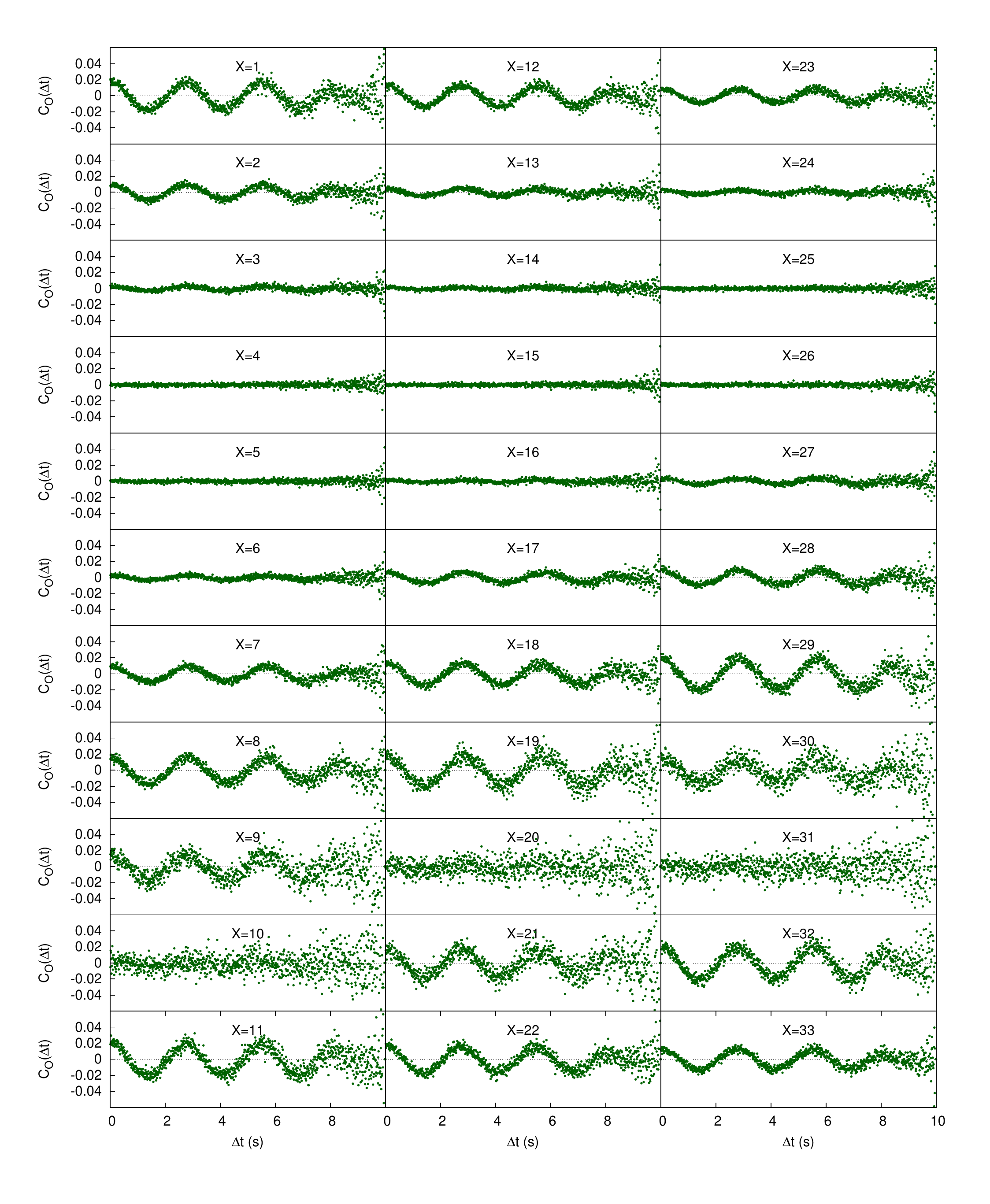} 
\caption{(color online) %
The correlation $C_\mathrm{O}(\Delta t)$ of the time-differences between the O-beam detection events for
different phase shifter settings $X$ as a function of the time difference $\Delta t$,
as obtained from the wave function collapse model with parameters
$A_\mathrm{O}=2745$,
$A_\mathrm{H}=4916$,
$B_\mathrm{O}=0.73$,
$B_\mathrm{H}=0.42$,
$\Omega_\mathrm{O}=0.60$,
$\chi_0=-2.7$,
$\epsilon_0=0.13$,
$Y=0.2$, $t_0=0$, and {\color{black}$\omega=2\pi/{\cal T}= 2\pi/2.8\;\mathrm{s}^{-1}$}.
}
\label{fig7}
\end{center}
\end{figure}

We scrutinize the ``one particle is one wave packet'' interpretation of quantum theory~\cite{RAUC15,BALL03} by
demonstrating that the standard way of invoking Born's postulate and the wave function collapse
may be reconciled with a description in terms of detection events and this
without invoking the notion of instantaneous action on a distance,
but only at the expense of attaching an extra attribute to the neutron.
The arguments that are used to construct the model go as follows:

\begin{enumerate}[(1)]
\item
We assume that each neutron is accompanied by an ``amplitude wave'' that is propagating through space.
At the beam splitters BS0, ..., BS3 this amplitude wave is transmitted and reflected
according to the rules of quantum theory.
\item
We assume that it is possible to identify spatially localized
``partial wave packets'' that represent the probability for observing a single neutron in the different beams.
\item
In line with (2), we assume that
the partial wave that appears in the O(H)-beam determines the probability that the detector in the O(H)-beam fires.
\item
The problem with the assumption in (3) is that there is a non-zero probability
that both the detectors in the O- and H-beam fire simultaneously (meaning within the time resolution
of the detection apparatus).

For instance, for $X=6$, the empirical frequencies for
an O- or H-beam event are roughly equal to one half (see Fig.~\ref{fig2}),
implying that the frequency with which both detectors fire simultaneously is about 1/4.

This is in blatant contradiction to the experimental data
in which the frequency for observing a simultaneous O- and H-beam detection event
is about $0.000069$.

\item
To get around the problem mentioned in (4),
some interpretations of quantum theory invoke the elusive wave function collapse to avoid
the clash with observed facts.
We wrote ``elusive'' because after its introduction by Heisenberg in 1927,
nobody has come up with a viable, realizable cause-and-effect model of how
this wave function collapse is supposed to happen in the single-neutron experiments.
In contrast, the statistical (= ensemble) interpretation does not make statements about
single events and therefore it cannot be used to develop an event-based description~\cite{BALL03}.
\item
Reasoning under the assumption that it is allowed to treat partial wave packets as independent entities,
we attach a new attribute, called ``{\bf A}'',
to the wave packet of each incident single neutron that enters the interferometer.
\item
Whenever a wave packet splits in two or more parts (according to the rules of quantum theory),
the attribute {\bf A} is assigned to one of these parts according to the probability of that particular part.
Other parts do not get the attribute {\bf A}.
\item
The key for making this artifice perform the ``wave function collapse''
is in the additional assumption that a neutron detector can only fire if it is hit
by a wave packet that carries the attribute {\bf A}.
\item
As only one partial wave carries the attribute {\bf A}, only one detector can fire at any time, i.e.
this cause-and-effect mechanism implements the wave function collapse
without invoking the notion of instantaneous action on a distance.
\end{enumerate}

Simulating this wave function collapse model just outlined on a computer is very simple, namely
generate either an O- or H-beam detection event with probability
\begin{eqnarray}
\widetilde P_\mathrm{O}&=&\frac{A_\mathrm{O}}{A_\mathrm{O}+A_\mathrm{H}}
\bigg(
1+ B_\mathrm{O}
\cos  \widetilde \Omega(t)
\bigg)
,
\label{sec4a1}
\\ 
\widetilde P_\mathrm{H}&=&\frac{A_\mathrm{H}}{A_\mathrm{O}+A_\mathrm{H}}
\bigg(1 - \frac{B_\mathrm{O}A_\mathrm{O}}{A_\mathrm{H}}
\cos \widetilde \Omega(t)
\bigg)
,
\label{sec4a2}
\end{eqnarray}
respectively, where $\widetilde \Omega(t)$ is given by Eq.~(\ref{sec4a0}).
The parameters  $A_\mathrm{O}$, $A_\mathrm{H}$, $B_\mathrm{O}$, $\Omega_\mathrm{O}$, $\chi_0$,
and $\epsilon_0$ are fixed by fitting the counts and standard deviations (see section~\ref{section3}).

The simulation results of the counts (not shown) and standard deviations (not shown)
are very similar to those of the experiment.
In Fig.~\ref{fig7} we present the simulation data of the
correlation $C_\mathrm{O}(\Delta t)$ of the time-differences between the O-beam detection events.
Comparing Fig.~\ref{fig7} with Fig.~\ref{fig5} suggests that the wave function collapse
model based on Eqs.~(\ref{sec4a1}) and (\ref{sec4a2}) reproduces the experimental findings quite well.
When present, the oscillations in the correlation $C_\mathrm{O}(\Delta t)$
persist for longer times than in the case of the real experiment but that
is to be expected because the wave function collapse model does not account for
sources of disturbances other than the oscillating phase difference.
Note that for fixed values of $\widetilde P_\mathrm{O}$ and $\widetilde P_\mathrm{H}$, this model
generates statistically independent detection events, the oscillations observed in the correlations
being due to the variation of $\widetilde P_\mathrm{O}$ and $\widetilde P_\mathrm{H}$ with time only.

\subsection{Discrete-event simulation}\label{section5b}

In the discrete-event simulation (DES) approach introduced in Ref.~\onlinecite{RAED05b},
there is no wave accompanying the particle.
Instead, interference results from the adaptive, cause-and-effect dynamics of
the model of a beam splitter.
In short, the application of DES to the single-neutron experiments is based on the following ideas:

\begin{figure}[t]
\begin{center}
\includegraphics[width=0.90\hsize]{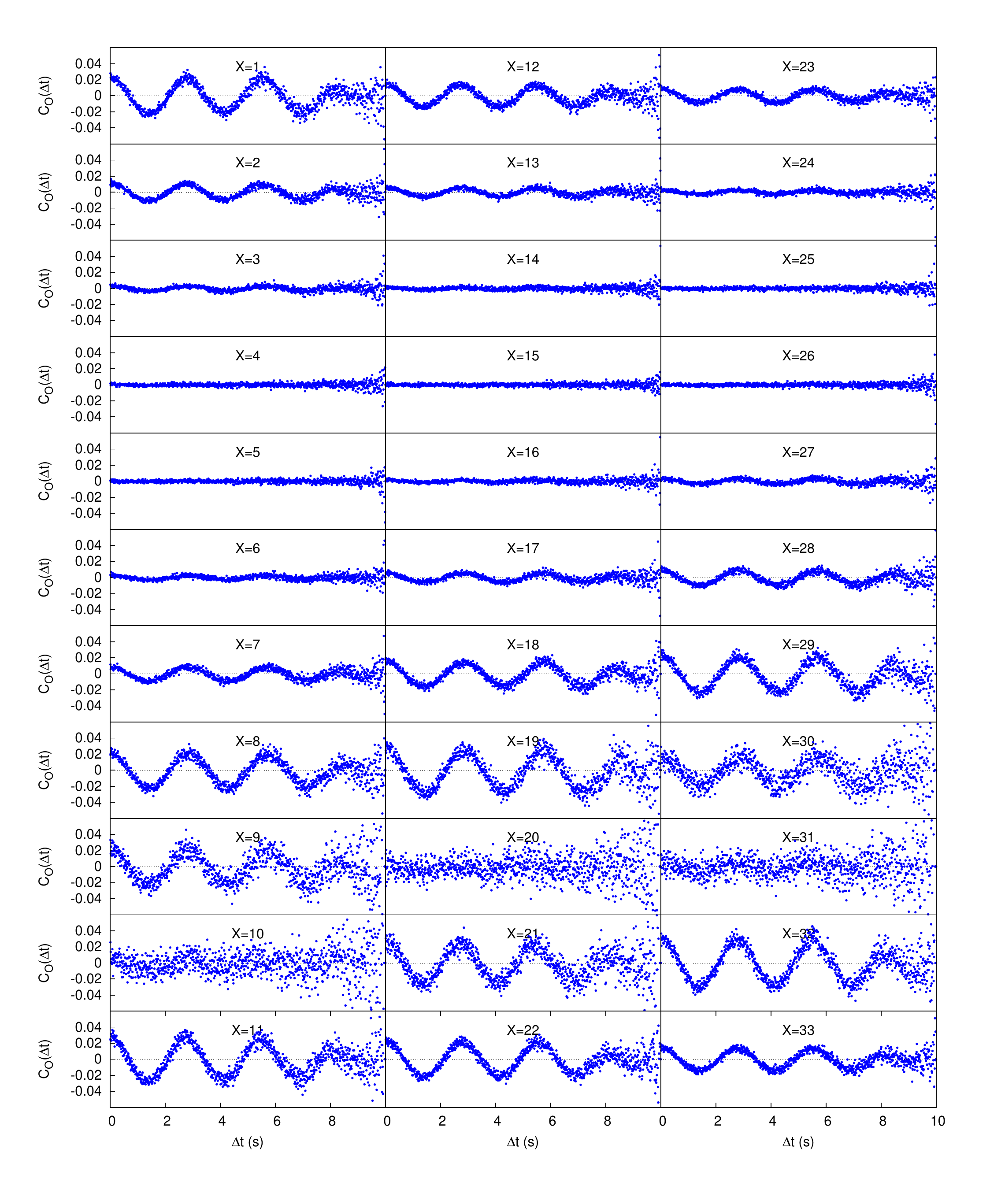} 
\caption{(color online) %
The correlation $C_\mathrm{O}(\Delta t)$ of the time-differences between the O-beam detection events for
different phase shifter settings $X$ as a function of the time difference $\Delta t$,
as obtained from a discrete-event simulation of the idealized neutron interference experiment
with an oscillating phase difference characterized by the parameters
$R=0.24$, $\gamma=0.6$ (parameters controlling the adaptive dynamics~\cite{RAED12b,RAED12a,MICH14a} of each beam splitter),
$\Omega_\mathrm{O}=0.60$,
$\chi_0=-2.7$,
$\epsilon_0=0.13$,
$Y=0.2$, $t_0=0$, and {\color{black}$\omega=2\pi/{\cal T}= 2\pi/2.8\;\mathrm{s}^{-1}$}.
}
\label{fig8}
\end{center}
\end{figure}

\begin{enumerate}[(1)]
\item
We treat a neutron as a single ``particle'', i.e. we do not need to introduce the concept of a wave
function and its associated consequences for the measurement of the neutron, that is the wave function collapse.
\item
Each neutron is processed one-by-one such that there is no direct interaction between the simulated neutrons.
\item
As the neutrons are simulated as particles, a neutron can only take one path through the interferometer.
\item
Each event registered by the O- or H-beam detector is caused by exactly one neutron entering
the interferometer. By construction, it is impossible to have one neutron cause the O- and H-beam detectors
to click simultaneously.
\item
Single-neutron interference is the result of the adaptive dynamics of the beam splitter model.
The adaptive dynamics itself is determined by the reflectivity $R$ and a parameter $\gamma$
which controls the speed of adaptation.
Input to the beam splitter model are the path on which the neutron enters and the time-of-flight
of the neutron.
The adaptive dynamics changes the internal state of the beam splitter model in a deterministic
manner.
The internal state determines the frequencies with which a neutron
randomly leaves the beam splitter in either the O- or in the H-beam.
\item
At no point in the formulation of the model, explicit use is made of
expressions like Eqs.~(\ref{sec4a1}) and ~(\ref{sec4a2}).
The input to the adaptive dynamic process results in updates of the internal state of the last beam splitter BS3.
This state causes the neutrons to leave BS3 in the O- or H-beam with frequencies that agree with
Eqs.~(\ref{sec4a1}) and ~(\ref{sec4a2}).
\end{enumerate}
For a detailed account of the implementation of the DES and its applications
to single-neutron experiments, see Refs.~\onlinecite{RAED12b,RAED12a,MICH14a}.

The DES reproduces the data of the counts (data not shown) and
of the standard deviation (data not shown) as a function of the setting $X$.
DES results of the correlation $C_\mathrm{O}(\Delta t)$ are shown in Fig.~\ref{fig8}.
Clearly, for all $X$, the DES results for $C_\mathrm{O}(\Delta t)$  compare
well with the corresponding entries in Fig.~\ref{fig5}.
When visible, the oscillations in the correlation $C_\mathrm{O}(\Delta t)$
persist for longer times than in the case of the real experiment but that
is to be expected because the DES does not account for
sources of disturbances other than the oscillating phase difference.

\begin{figure}[t]
\begin{center}
\includegraphics[width=0.90\hsize]{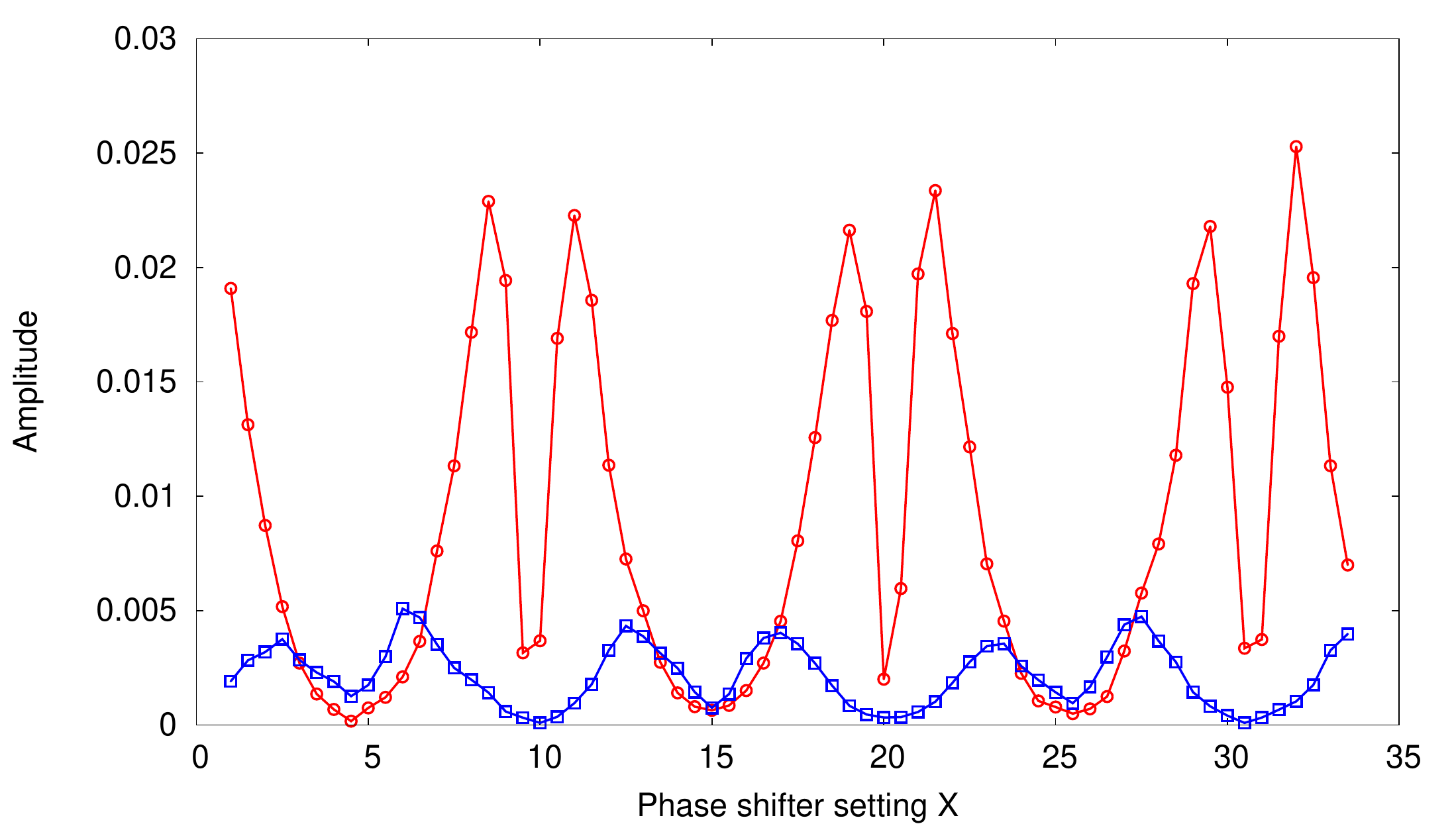} 
\caption{(color online) %
The maximum amplitudes of the oscillations in
$C_{\mathrm{O}}(\Delta t)$ ($\bigcirc$)
and
$C_{\mathrm{H}}(\Delta t)$ ($\Box$),
computed from the DES data,
as a function of the phase shifter setting $X$.
Lines through the data points are guides to the eye.
}
\label{fig8z}
\end{center}
\end{figure}

Figure~\ref{fig8z} shows the maximum amplitudes of the oscillations in $C_{\mathrm{O}}(\Delta t)$ and
$C_{\mathrm{H}}(\Delta t)$ as a function of the phase shifter setting $X$, as obtained from the DES.
Comparing Fig.~\ref{fig5z} with Fig.~\ref{fig8z}, we conclude that DES reproduces the
main features of the oscillations found in the experimental data.

Note that also in the DES model, the detection events are not statistically independent,
the oscillations observed in the correlations being due to the variation of the path length with time.
Obviously, the neutron data itself do not contain enough information to identify the source of these variations
with certainty but plausible causes are small vibrations of the aluminum plate that is used for the phase shifter
or small internal vibrations of the interferometer (i.e. the silicon single-crystal) itself,
due to temperature or other environmental fluctuations.

\section{Discussion}\label{section6}

In this paper, we have presented a detailed analysis of the time series
of time-stamped neutron counts obtained by single-neutron interferometry,
the results of which can be summarized as follows:
\begin{enumerate}[(1)]
\item
The neutron counting statistics displays the Poissonian behavior as expected~\cite{RAUC15}.
\item
The dependence of the variance of the neutron counts on the phase shifter setting
can be explained by a probabilistic model that accounts for fluctuations of the phase shift.
\item
The time series of the detection events exhibit long-time correlations,
the amplitude of which depends on the phase shifter setting.
These correlations appear as damped oscillations
with a period of about {\color{black}2.8} seconds, a fairly long time relative to the average
time of 1.3 milliseconds between the detection of two successive neutrons.
The cause of this oscillations is unknown.
Viable candidate causes are fluctuations induced by temperature and/or vibrations.

\item
We have shown by simulation that the correlations of the time differences observed in experiment
can be reproduced by assuming that for a fixed setting of the phase shifter,
the phase shift experienced by the neutrons varies periodically in time with a period of
{\color{black}${\cal T}\approx2.8\;\mathrm{s}$}.
\end{enumerate}

The results presented in this paper demonstrate that time-stamped data of
single-particle interference experiments may exhibit features that go beyond the standard
quantum model of independent random events and suggest instead a description in terms of non-stationary processes.
With the available data, it is not possible to identify a cause for the observed correlations.
It would be of interest to repeat the experiments with substantially longer measuring
times {\color{black}and better statistics} to see if the oscillations persist over a much longer time interval, before they disappear in the statistical noise.
Another point to check might be the mechanical properties of the phase shifter, perhaps by using another one.
Finally, although we have found the oscillations to persist over a long period, it might be worthwhile
to repeat the same experiment, {\color{black} preferably at another neutron facility such as the one of NIST}.
{\color{black}
If such an experiment confirms that the correlations exhibit oscillations with a frequency of about 2.8 seconds, the implication is that this phenomenon is beyond the prediction of a theory of independent random events.
}
\section*{Acknowledgements}
We would like to thank Helmut Rauch, Koen De Raedt and Seiji Miyashita for helpful suggestions and discussions.
This work is dedicated to the memory of Helmut Rauch.

\bibliography{../../../all20}
\end{document}